\documentstyle[epsf]{mn2e}

\newcommand{\sD}{{\cal D}}
\newcommand{\sF}{{\cal F}}
\newcommand{\sG}{{\cal G}}

\newcommand{\bB}{\mbox{\boldmath$B$}}
\newcommand{\bE}{\mbox{\boldmath$E$}}
\newcommand{\bX}{\mbox{\boldmath$X$}}
\newcommand{\be}{\mbox{\boldmath$e$}}
\newcommand{\bk}{\mbox{\boldmath$k$}}
\newcommand{\bx}{\mbox{\boldmath$x$}}
\newcommand{\bu}{\mbox{\boldmath$u$}}
\newcommand{\bOmega}{\mbox{\boldmath$\Omega$}}

\title[Magnetorotational turbulent stresses]
{On the dynamics of magnetorotational turbulent stresses}

\author[G. I. Ogilvie]
  {G. I. Ogilvie$^{1,2}$\\
  $^1$Institute of Astronomy, University of Cambridge, Madingley Road,
  Cambridge CB3 0HA\\
  $^2$Department of Applied Mathematics and Theoretical Physics,
  University of Cambridge, Silver Street, Cambridge CB3 9EW}

\begin{document}

\maketitle

\label{firstpage}

\begin{abstract}
  The turbulent stresses that lead to angular momentum transport in
  accretion discs have often been treated as resulting from an
  isotropic effective viscosity, related to the pressure through the
  alpha parametrization of Shakura \& Sunyaev.  This simple approach
  may be adequate for the simplest aspects of accretion disc theory,
  and was necessitated historically by an incomplete understanding of
  the origin of the turbulence.  More recently, Balbus \& Hawley have
  shown that the magnetorotational instability provides a robust
  mechanism of generating turbulent Reynolds and Maxwell stresses in
  sufficiently ionized discs.  The alpha viscosity model fails to
  describe numerous aspects of this process.  This paper introduces a
  new analytical model that aims to represent more faithfully the
  dynamics of magnetorotational turbulent stresses and bridge the gap
  between analytical studies and numerical simulations.  Covariant
  evolutionary equations for the mean Reynolds and Maxwell tensors are
  presented, which correctly include the linear interaction with the
  mean flow.  Non-linear and dissipative effects, in the absence of an
  imposed magnetic flux and in the limit of large Reynolds number and
  magnetic Reynolds number, are modelled through five non-linear terms
  that represent known physical processes and are strongly constrained
  by symmetry properties and dimensional considerations.  The
  resulting model explains the development of statistically steady,
  anisotropic turbulent stresses in the shearing sheet, a local
  representation of a differentially rotating disc, in agreement with
  numerical simulations.  It also predicts that purely hydrodynamic
  turbulence is not sustained in a flow that adequately satisfies
  Rayleigh's stability criterion.  The model is usually formally
  hyperbolic and therefore `causal', and guarantees the realizability
  of the stress tensors.  It should be particularly useful in
  understanding the dynamics of warped, eccentric and tidally
  distorted discs, non-Keplerian accretion flows close to black holes,
  and a variety of time-dependent accretion phenomena.
\end{abstract}


\section{Introduction}

The magnetorotational instability is one of the most important
instabilities in astrophysical fluid dynamics.  It applies to a
differentially rotating, electrically conducting fluid in which the
angular velocity decreases in magnitude away from the axis of
rotation.\footnote{In the absence of entropy gradients.  Otherwise,
  modified versions of the H\o iland stability criteria exist
  (Papaloizou \& Szuszkiewicz 1992; Balbus 1995).}  In the presence of
a weak magnetic field of arbitrary configuration, such a flow is
subject to a dynamical instability, with a growth rate comparable to
the shear rate of the flow (Velikhov 1959; Chandrasekhar 1960; Fricke
1969; Acheson 1978; Balbus \& Hawley 1991, 1992; Papaloizou \&
Szuszkiewicz 1992; Balbus 1995; Foglizzo \& Tagger 1995; Ogilvie \&
Pringle 1996; Terquem \& Papaloizou 1996).

The principal application of the magnetorotational instability is to
accretion discs, in which the profile of angular velocity is fixed by
Kepler's third law.  The non-linear development of the instability
leads to sustained magnetohydrodynamic (MHD) turbulence, which
transports angular momentum outwards in a vain attempt to neutralize
the destabilizing gradient of angular velocity (Hawley, Gammie \&
Balbus 1995; Brandenburg et al. 1995; Stone et al. 1996; Balbus \&
Hawley 1998).  Owing to the generality of the conditions for
instability, however, further applications exist to stellar interiors
and other astrophysical objects, and possibly to geophysical
situations.  Laboratory experiments are also in progress (Ji, Goodman
\& Kageyama 2001).

In one sense, the magnetorotational instability is generally
considered to have `solved' the problem of the effective viscosity of
accretion discs, through providing a robust means of angular momentum
transport.\footnote{Nevertheless, considerable interest and debate
  remain regarding the operation of the magnetorotational instability,
  or alternative mechanisms of angular momentum transport, in weakly
  ionized discs, e.g. around young stars and in the quiescent state of
  dwarf novae.}  However, the realization of the importance and
ubiquity of the instability has, in some ways, made matters more
difficult.  A strict view might be taken that any calculation of an
accretion disc must now involve a numerical simulation of
three-dimensional MHD turbulence.  These simulations are demanding to
execute, complicated to analyse and still restricted in their scope
and physical content.

To illustrate the limitations of the direct numerical approach, let us
consider how the computational requirements scale with the parameters
of the problem being studied.  Most simulations conducted so far have
been based on the shearing box (Hawley et al. 1995), which is a local
representation of a differentially rotating disc.  The simulated
volume is comparable to $H^3$, where $H$ is the vertical scale-height
or semi-thickness of the disc, and the models can be run for many
times the dynamical (orbital) time-scale.  Let us accept, for the
purposes of this argument, that recent versions of such simulations
(e.g. Miller \& Stone 2000) have achieved an adequate fidelity to
physical reality (although questions indeed remain concerning
numerical resolution and convergence, vertical boundary conditions,
thermal and radiative physics, etc.).  How much more difficult would
it be to simulate a global model of a realistic thin disc over many
times the viscous time-scale, as would be required in a typical
application?  Suppose that the disc has a constant angular
semi-thickness $H/r$ and is simulated in a conical wedge in spherical
polar coordinates $(r,\theta,\phi)$ with logarithmic grid spacing in
$r$ between $r_{\rm in}$ and $r_{\rm out}$, and uniform spacing in
$\theta$ and $\phi$.  To achieve a resolution comparable to that of a
shearing box, the number of grid zones must increase by a factor of
order $(r/H)\ln(r_{\rm out}/r_{\rm in})$ in $r$, and a factor of order
$r/H$ in $\phi$.  The global viscous time-scale exceeds the dynamical
time-scale in the inner part of the disc (where the Courant condition
on the time-step is most severe) by a factor of order
$(1/\alpha)(r/H)^2(r_{\rm out}/r_{\rm in})^{3/2}$, where $\alpha$ is
the viscosity parameter of Shakura \& Sunyaev (1973).  Therefore the
computational requirements increase by a factor of order
\begin{equation}
  {{1}\over{\alpha}}\left({{r}\over{H}}\right)^4
  \left({{r_{\rm out}}\over{r_{\rm in}}}\right)^{3/2}
  \ln\left({{r_{\rm out}}\over{r_{\rm in}}}\right).
\end{equation}
Reasonable estimates of this factor for discs around young stars, in
cataclysmic variables, and in X-ray binaries, respectively, are
$10^{13}$, $10^{11}$ and $10^{15}$, or even larger.  (These numbers
are reduced if the inner part of the disc is truncated by a stellar
magnetosphere.)  For this reason, although some aspects of large-scale
disc dynamics can now be addressed in direct numerical simulations
(e.g.  Hawley 2001), global studies of realistic thin discs are
currently inaccessible.

How then are we to study situations such as the thermal-viscous
instability of discs in cataclysmic variables, the dynamics of warped
or eccentric discs, tidally distorted discs in binary stars and
protoplanetary systems, non-Keplerian accretion flows close to black
holes, or the propagation of waves in discs?  The traditional
approach, of course, has been to treat the turbulence as an effective
viscosity given by the alpha parametrization of Shakura \& Sunyaev
(1973).  Despite suspicions that this treatment is likely to be
inadequate in various respects, very few studies have aimed
specifically at testing this issue (Abramowicz, Brandenburg \& Lasota
1996; Torkelsson et al. 2000) and it is still unclear in what respects
magnetorotational turbulence does or does not behave as an effective
viscosity.  In a standard accretion disc the rate-of-strain tensor has
a dominant $r\phi$-component and an isotropic viscous model predicts
that the $r\phi$-component of the turbulent stress would similarly
dominate.  This is not true of magnetorotational turbulence, in which
other stress components (e.g. $\phi\phi$) are larger.  However, for
the simplest problems such as the evolution of the surface density of
a standard accretion disc, these additional stress components play
essentially no role in the dynamics.  The alpha model is probably
perfectly adequate in these circumstances, where one dimensionless
number, $\alpha$, is parametrizing a single physical quantity, the
vertically integrated stress coefficient $T_{r\phi}$.\footnote{It may
  not be adequate for predicting the vertical distribution of energy
  dissipation, and therefore the vertical structure of the disc.}

In situations such as warped, eccentric, tidally distorted and
non-Keplerian discs there is a large-scale, possibly slowly varying
velocity field that is different from that in a standard disc.  From
the perspective of a local observer, however, the flow resembles a
shearing-box model, but with a non-standard (and possibly slowly
varying) velocity gradient tensor.  Stress components other than the
$r\phi$-component can play a significant role in the dynamics of these
systems, but whether the full turbulent stress tensor resembles a
viscous stress in such situations is not understood.  In principle,
shearing-box studies could be used to formulate a more faithful
analytical or semi-analytical model of the relevant properties of the
turbulence, which could then be used in studies of the large-scale
behaviour of accretion discs.  This would bridge the gap between the
local dynamics (on scales from the dissipation length up to $H$) and
the large-scale dynamics, without requiring colossal advances in
computation.  The formation and analysis of such a model is also
likely to result in a better physical understanding.

In view of the formidable complexity of turbulent flows, such a
programme might appear overambitious.  A complete theory of turbulence
does not exist, because turbulent flows have an unlimited number of
statistical properties that cannot be calculated from first
principles.  However, in the present context, our attention is
restricted to the question of how the mean turbulent stress in a patch
of the disc responds to changes in the large-scale velocity gradient.
As will be made clear, certain linear aspects of this problem can be
deduced directly from the MHD equations; other non-linear aspects
cannot be treated rigorously and require a closure model that is,
nevertheless, strongly constrained by symmetry properties and
dimensional considerations.  Although the model affords only an
imperfect description of turbulence, its physically motivated nature
and its connection to the MHD equations ensure that it performs more
faithfully than the alpha viscosity model, and this is borne out by
numerous test problems.

The remainder of this paper is organized as follows.  In
Section~\ref{A minimal model system for magnetorotational turbulence}
I introduce a conceptual model system for magnetorotational
turbulence, and make some simple arguments concerning the saturation
of the turbulence.  I develop the equations for the mean Reynolds and
Maxwell tensors as far as possible analytically in Section~\ref{Stress
  equations and closures}, and then discuss the requirements of a
non-linear closure model.  A specific model satisfying these
principles is presented and explained physically in Section~\ref{A
  simple model and its properties}.  The outcome of the model in the
shearing sheet, under a variety of conditions, is investigated in
Section~\ref{Non-linear outcome in the shearing sheet}.  In
Section~\ref{The governing equations in compressible flows} the
effects of a compressible mean flow are incorporated and the total
energy budget is explained.  Some preliminary applications of the
model are worked out in Section~\ref{Applications and implications}.
In Section~\ref{Comparison with previous work} a comparison is made
with existing models in engineering and in astrophysics, and also with
published numerical simulations.  A summary follows in
Section~\ref{Conclusion}.

\section{A minimal model system for magnetorotational turbulence}

\label{A minimal model system for magnetorotational turbulence}

The magnetorotational instability has certain minimal requirements:
rotation and shear (in the correct relative orientation), electrical
conductivity and a seed magnetic field.  Let us consider the simplest
conceptual model system for studying magnetorotational turbulence: the
incompressible shearing sheet (Goldreich \& Lynden-Bell 1965).  Let
$(x,y,z)$ be Cartesian coordinates in a frame of reference rotating
with uniform angular velocity $\bOmega=\Omega\,\be_z$.  An
incompressible fluid of uniform density $\rho$, kinematic viscosity
$\nu$ and magnetic diffusivity $\eta$ has a uniformly shearing
velocity field characterized by a uniform velocity gradient tensor
$\nabla\bu$.  In a standard disc this is of the form
$\bu=-2Ax\,\be_y$, where $A$ is Oort's first constant.  The $x$, $y$
and $z$ directions correspond to the radial, azimuthal and vertical
directions from the perspective of a corotating observer in a thin,
differentially rotating disc.  The flow is unbounded in the horizontal
($xy$) plane but of bounded vertical extent, with either `physical' or
periodic boundary conditions.  The distance between the vertical
boundaries defines a characteristic length-scale $L=L_z$ which, in a
real disc, is related to the vertical scale-height or thickness.

As defined above, the system involves just three dimensionless
parameters: the Rossby number ${\rm Ro}=A/\Omega$, the Reynolds number
${\rm Re}=L^2\Omega/\nu$ and the magnetic Prandtl number ${\rm
  Pm}=\nu/\eta$.  In a Keplerian disc, the Rossby number is fixed at
${\rm Ro}=3/4$.  The microscopic diffusivities $\nu$ and $\eta$ are
included in this conceptual model only to regularize the system by
allowing for dissipation and irreversibility.  The Reynolds number may
be considered to be arbitrarily large.

Apart from the assumption of incompressibility and the inclusion of
microscopic diffusivities, this system is equivalent to the shearing
box studied by Hawley et al. (1995) in the limit that the horizontal
scales $L_x$ and $L_y$ of the box tend to infinity while retaining a
finite vertical scale.  Note that the limit $L_x, L_y\gg L_z$
corresponds to the physical situation in a thin disc.  It is natural
to make the following assumptions, which are not contradicted by
existing numerical simulations:
\begin{itemize}
\item the macroscopic statistical properties of the turbulence do not
  depend on the details of the dissipative mechanisms, and are
  therefore independent of ${\rm Re}$ and ${\rm Pm}$ in the limit
  ${\rm Re}\to\infty$;
\item the macroscopic statistical properties of the turbulence are
  bounded and well defined in the limit $L_x,L_y\to\infty$.
\end{itemize}

The system is horizontally homogeneous, in the sense that any point in
the $xy$-plane is equivalent (modulo a Galilean boost).  If periodic
vertical boundary conditions are imposed, the system is also
vertically homogeneous.  It is then possible for the system to develop
statistically steady and homogeneous (although anisotropic)
turbulence.

If magnetorotational turbulence is initiated in this system, the
properties of the saturated state are strongly constrained by
dimensional considerations.  The RMS turbulent velocity fluctuation,
$\langle u^2\rangle^{1/2}$, for example, must equal $L\Omega$ times a
dimensionless coefficient of order unity, if the assumed independence
of ${\rm Re}$ and ${\rm Pm}$ is correct.  Therefore the vertical scale
of the system plays an essential role in the non-linear saturation of
the turbulence, presumably by limiting the size of coherent structures
(`eddies').  In an unbounded system with $L\to\infty$, the turbulence
would presumably grow indefinitely without saturation.  It is
sometimes suggested that turbulent motions and magnetic fields in an
accretion disc are limited by shock formation or magnetic buoyancy,
but there is little evidence of this in numerical simulations of
magnetorotational turbulence, and neither limiting mechanism operates
in the incompressible shearing sheet.\footnote{Numerical simulations
  of an incompressible shearing box using a spectral method
  (unpublished work by the author) produce results similar to those of
  Hawley et al. (1995).}

\section{Stress equations and closures}

\label{Stress equations and closures}

\subsection{Basic equations}

Consider an incompressible fluid of uniform density $\rho$, kinematic
viscosity $\nu$ and magnetic diffusivity $\eta$.  In the Cartesian tensor
notation, the equation of motion in a frame of reference rotating with
uniform angular velocity $\Omega_i$ is
\begin{equation}
  (\partial_t+u_j\partial_j)u_i+2\epsilon_{ijk}\Omega_ju_k=
  -\partial_i\Pi+b_j\partial_jb_i+\nu\partial_{jj}u_i,
\end{equation}
where
\begin{equation}
  b_i=(\mu_0\rho)^{-1/2}B_i
\end{equation}
is the Alfv\'en velocity and
\begin{equation}
  \Pi={{p}\over{\rho}}+{\textstyle{{1}\over{2}}}b^2+\Phi
\end{equation}
is the modified pressure including the gas pressure, the magnetic
pressure and the gravitational and centrifugal potentials.  The
induction equation is
\begin{equation}
  (\partial_t+u_j\partial_j)b_i=b_j\partial_ju_i+\eta\partial_{jj}b_i,
\end{equation}
and the velocity and magnetic fields satisfy the solenoidal conditions,
\begin{equation}
  \partial_iu_i=\partial_ib_i=0.
\end{equation}

Separate the varying quantities into mean and fluctuating parts, e.g.
\begin{equation}
  u_i=\bar u_i+u_i',
\end{equation}
where $\bar u_i=\langle u_i\rangle$ and $\langle u_i'\rangle=0$.  The
angle brackets denote a suitable averaging procedure such as a
spatial, temporal or ensemble average.  The mean forms of the above
equations are then
\begin{eqnarray}
  \lefteqn{(\partial_t+\bar u_j\partial_j)\bar u_i+
  2\epsilon_{ijk}\Omega_j\bar u_k=
  -\partial_i\bar\Pi+\bar b_j\partial_j\bar b_i+\nu\partial_{jj}\bar u_i}
  &\nonumber\\
  &&+\partial_j(\bar M_{ij}-\bar R_{ij}),
\end{eqnarray}
\begin{equation}
  (\partial_t+\bar u_j\partial_j)\bar b_i=\bar b_j\partial_j\bar u_i+
  \eta\partial_{jj}\bar b_i+\partial_j\bar F_{ij},
\end{equation}
\begin{equation}
  \partial_i\bar u_i=\partial_i\bar b_i=0,
\end{equation}
where
\begin{equation}
  M_{ij}=b_i'b_j',
\end{equation}
\begin{equation}
  R_{ij}=u_i'u_j',
\end{equation}
\begin{equation}
  F_{ij}=u_i'b_j'-u_j'b_i'
\end{equation}
are the Maxwell, Reynolds and Faraday tensors.  The problem at hand is
how to determine the mean quantities $\bar M_{ij}$, $\bar R_{ij}$ and
$\bar F_{ij}$ and thereby close the system of equations.
\footnote{For the high-Reynolds number situations considered in this
  paper, the diffusive terms involving $\nu$ and $\eta$ in the mean
  equations may be neglected.}

Unless the mean velocity field acts as a dynamo, the mean magnetic
field can be sustained only by the mean Faraday tensor, which
represents the turbulent electromotive force or `alpha effect'.  For
simplicity, it will be assumed in the following that there is no mean
magnetic field and no large-scale dynamo, i.e. $\bar b_i=\bar
F_{ij}=0$.  The remaining problem is to determine the Maxwell and
Reynolds stress tensors that appear in the mean equation of motion.

In the case $\bar b_i=\bar F_{ij}=0$, the induction equation reads
\begin{equation}
  (\partial_t+\bar u_j\partial_j)b_i'+u_j'\partial_jb_i'=
  b_j'\partial_j\bar u_i+b_j'\partial_ju_i'+\eta\partial_{jj}\bar b_i'.
\end{equation}
From this, an exact equation for the mean Maxwell stress can be
obtained, in the form
\begin{eqnarray}
  \lefteqn{(\partial_t+\bar u_k\partial_k)\bar M_{ij}-
  \bar M_{ik}\partial_k\bar u_j-\bar M_{jk}\partial_k\bar u_i}&\nonumber\\
  &&=\langle b_i'\partial_kF_{jk}+b_j'\partial_kF_{ik}\rangle+
  \eta\langle b_i'\partial_{kk}b_j'+b_j'\partial_{kk}b_i'\rangle.
\end{eqnarray}
The left-hand side of this equation represents the linear dynamics of
the Maxwell tensor.  It shows how the tensor is advected by the mean
velocity field and `stretched' by gradients in the mean velocity.  The
right-hand side contains difficult terms of two types: those involving
triple correlations of fluctuating quantities, and those involving the
microscopic diffusion process.  These are both `non-linear' effects;
although Ohmic diffusion is a linear process, when the magnetic
Reynolds number is large the diffusive terms can be significant only
when a turbulent cascade has forced structure to appear on the
dissipative scales.

There is little hope of finding a rigorous closure scheme for this
equation.  Instead, the approach adopted here will be to write the
equation in the form
\begin{equation}
  (\partial_t+\bar u_k\partial_k)\bar M_{ij}-
  \bar M_{ik}\partial_k\bar u_j-\bar M_{jk}\partial_k\bar u_i=\cdots,
  \label{mij_linear}
\end{equation}
retaining the exact form of the linear terms, and then investigate
simple closure models for the non-linear terms on the right-hand side.

A similar approach applied to the fluctuating part of the equation of
motion leads to the exact equation
\begin{eqnarray}
  \lefteqn{(\partial_t+\bar u_k\partial_k)\bar R_{ij}+
  \bar R_{ik}\partial_k\bar u_j+\bar R_{jk}\partial_k\bar u_i}&\nonumber\\
  &&+2\epsilon_{jkl}\Omega_k\bar R_{il}+2\epsilon_{ikl}\Omega_k\bar R_{jl}=
  -\langle u_i'\partial_j\Pi'+u_j'\partial_i\Pi'\rangle\nonumber\\
  &&+\langle u_i'\partial_k(M_{jk}-R_{jk})+
  u_j'\partial_k(M_{ik}-R_{ik})\rangle\nonumber\\
  &&+\nu\langle u_i'\partial_{kk}u_j'+u_j'\partial_{kk}u_i'\rangle
  \label{rij_exact}
\end{eqnarray}
for the mean Reynolds tensor.  The terms on the right-hand side
require a closure model, but the form of the linear terms,
\begin{eqnarray}
  \lefteqn{(\partial_t+\bar u_k\partial_k)\bar R_{ij}+
  \bar R_{ik}\partial_k\bar u_j+\bar R_{jk}\partial_k\bar u_i}&\nonumber\\
  &&+2\epsilon_{jkl}\Omega_k\bar R_{il}+2\epsilon_{ikl}\Omega_k\bar R_{jl}=
  \cdots,
  \label{rij_linear}
\end{eqnarray}
may be retained.  Note that the Maxwell and Reynolds tensors interact
with the mean velocity gradient in different ways, and only the
Reynolds tensor is affected by rotation.

\subsection{The pressure--strain correlation}

In homogeneous turbulence, the term involving $\Pi'$ on the right-hand
side of equation (\ref{rij_exact}) has the alternative form
\begin{equation}
  \langle\Pi'(\partial_ju_i'+\partial_iu_j')\rangle,
\end{equation}
and is known as the pressure--strain correlation.  The identification
of suitable closures for this term, even in purely hydrodynamic
turbulence, has been a matter of considerable controversy (e.g.
Speziale 1991) leading to highly elaborate, but still imperfect,
models (e.g. Sj\"ogren \& Johansson 2000).  An expression for $\Pi'$
can be obtained by taking the divergence of the fluctuating part of
the equation of motion.  If the spectrum of the turbulence is known,
part of the pressure--strain correlation can then be expressed in
terms of $\bar R_{ij}$, while part is undoubtedly non-linear.
However, as the spectrum itself is determined through non-linear
dynamics, one may take the view that the entire pressure--strain
correlation is a non-linear term for which a non-deductive closure
must be proposed.

In the following, the fluctuating quantities will not be referred to,
and the bars on $\bar u_i$, $\bar M_{ij}$ and $\bar R_{ij}$ will be
omitted.

\subsection{Linear dynamics in the shearing sheet}

\label{linear}

In the shearing sheet the angular velocity is
${\bf\Omega}=\Omega\,\be_z$ and the only non-vanishing component of the
mean velocity gradient is $\partial_xu_y=-2A$.  The linearized
equations for the Maxwell and Reynolds stresses then have the form
\begin{eqnarray}
  &&\partial_tM_{xx}=0,\nonumber\\
  &&\partial_tM_{xy}+2AM_{xx}=0,\nonumber\\
  &&\partial_tM_{xz}=0,\nonumber\\
  &&\partial_tM_{yy}+4AM_{xy}=0,\nonumber\\
  &&\partial_tM_{yz}+2AM_{xz}=0,\nonumber\\
  &&\partial_tM_{zz}=0,\nonumber\\
  &&\partial_tR_{xx}-4\Omega R_{xy}=0,\nonumber\\
  &&\partial_tR_{xy}+2(\Omega-A)R_{xx}-2\Omega R_{yy}=0,\nonumber\\
  &&\partial_tR_{xz}-2\Omega R_{yz}=0,\nonumber\\
  &&\partial_tR_{yy}+4(\Omega-A)R_{xy}=0,\nonumber\\
  &&\partial_tR_{yz}+2(\Omega-A)R_{xz}=0,\nonumber\\
  &&\partial_tR_{zz}=0.\nonumber
\end{eqnarray}
The Maxwell and Reynolds tensors are decoupled.  The general solution
for the magnetic component is
\begin{eqnarray}
  &&M_{xx}=M_{xx0},\nonumber\\
  &&M_{xy}=M_{xy0}-2M_{xx0}At,\nonumber\\
  &&M_{xz}=M_{xz0},\nonumber\\
  &&M_{yy}=M_{yy0}-4M_{xy0}At+4M_{xx0}A^2t^2,\nonumber\\
  &&M_{yz}=M_{yz0}-2M_{xz0}At,\nonumber\\
  &&M_{zz}=M_{zz0},\nonumber
\end{eqnarray}
and allows for algebraic growth through the shearing of magnetic field
lines.  The solution for the kinetic component depends on the Rayleigh
discriminant, or squared epicyclic frequency,
\begin{equation}
  \kappa^2=4\Omega(\Omega-A),
\end{equation}
of the system.  In the absence of rotation, the system is
Rayleigh-neutral and the solution for $R_{ij}$ allows for algebraic
growth (it resembles the solution for $M_{ij}$, but with the sign of
$A$ reversed).  When $\kappa^2>0$, the solution is oscillatory, as can
be seen by reducing the problem to equations of the form
\begin{eqnarray}
  &&\partial_{tt}R_{xy}+4\kappa^2R_{xy}=0,\nonumber\\
  &&\partial_{tt}R_{xz}+2\kappa^2R_{xz}=0,\nonumber\\
  &&\partial_{tt}R_{zz}=0.\nonumber
\end{eqnarray}
So $R_{xx}$, $R_{xy}$ and $R_{yy}$ oscillate at twice the epicyclic
frequency, $R_{xz}$ and $R_{yz}$ at the epicyclic frequency, and
$R_{zz}$ is constant.  On the other hand, when $\kappa^2<0$,
Rayleigh's criterion for stability is violated and exponential growth
occurs.

\subsection{Requirements of a non-linear closure model}

Whenever the linear dynamics indicates unbounded growth or undamped
oscillations, the non-linear terms will determine the eventual
outcome.  The approach adopted in this paper is to explore the
consequences of simple closure models of the form
\begin{eqnarray}
  \sD^{(1)}R_{ij}&=&\sF_{ij}^{(1)}(R_{ij},M_{ij},\dots),\nonumber\\
  \sD^{(2)}M_{ij}&=&\sF_{ij}^{(2)}(R_{ij},M_{ij},\dots),
\end{eqnarray}
where the operators $\sD$ are defined by the left-hand sides of
equations (\ref{rij_linear}) and (\ref{mij_linear}) respectively, and
the quantities $\sF_{ij}$ are non-linear tensorial functions of their
arguments.  The dots represent the parameters of the problem, on which
the functions $\sF_{ij}$ may depend.

Such a model ought to satisfy the following fundamental principles.
\begin{itemize}
\item There should be no `source terms'.  $R_{ij}=M_{ij}=0$ should
  always be a solution, although it may be unstable.
\item An unmagnetized state, $M_{ij}=0$, should always be a solution
  even when $R_{ij}\ne0$, although it may be unstable to dynamo
  action.
\item The non-linear terms on the right-hand side should be manifestly
  covariant, as the left-hand sides are.
\item The positive semi-definite nature of the tensors $R_{ij}$ and
  $M_{ij}$, implicit in their definition, should be preserved by the
  model.\footnote{A real symmetric $n\times n$ matrix (or second-rank
    tensor) $A_{ij}$ is said to be positive semi-definite if
    $A_{ij}X_iX_j\ge0$ for all real $n$-component vectors $X_i$, or,
    equivalently, if all the eigenvalues of $A_{ij}$ are non-negative.
    This condition is relevant because $\bar R_{ij}X_iX_j=\langle
    u_i'u_j'X_iX_j\rangle=\langle(\bu'\cdot\bX)^2\rangle\ge0$, and
    similarly for $\bar M_{ij}$.  Note that off-diagonal components
    such as $\bar R_{xy}$ can have either sign.}  This ensures not
  only that the turbulent kinetic and magnetic energy densities remain
  non-negative, but also that the stress tensors can be realized by
  genuine velocity and magnetic fields.
\end{itemize}
The following two additional requirements appear plausible, although
they cannot be regarded as fundamental principles.
\begin{itemize}
\item The non-linear terms should not refer to the angular velocity
  $\bf\Omega$ or the velocity gradient $\nabla\bu$, because their
  effects are fully represented in the linear terms.
\item The non-linear terms should not refer to the microscopic
  diffusion coefficients, because one has assumed asymptotic
  independence of ${\rm Re}$ and ${\rm Pm}$ in the limit ${\rm
    Re}\to\infty$.
\end{itemize}
The non-linear terms are also strongly constrained by dimensional
considerations.  Both $R_{ij}$ and $M_{ij}$, as defined in this
incompressible system, have the dimensions of velocity-squared.  A
quantity with the dimensions of the rate of change of $R_{ij}$ cannot
be formed from $R_{ij}$ and $M_{ij}$ alone.  The only other physical
quantity that can appear in the functions $\sF$, according to the
above assumptions, is the length-scale $L$.  In that case the
non-linear terms must have the form
\begin{equation}
  \sF_{ij}^{(n)}=L^{-1}\sG_{ij}^{(n)}(R_{ij},M_{ij}),
\end{equation}
where the quantities $\sG_{ij}$ are homogeneous functions of degree
$3/2$.

\section{A simple model and its properties}

\label{A simple model and its properties}

\subsection{Statement of the model}

A simple model based on the above principles is as follows.
\begin{eqnarray}
  \lefteqn{(\partial_t+u_k\partial_k)R_{ij}+
  R_{ik}\partial_ku_j+R_{jk}\partial_ku_i}&\nonumber\\
  &&\qquad+2\epsilon_{jkl}\Omega_kR_{il}+2\epsilon_{ikl}\Omega_kR_{jl}
  \nonumber\\
  &&=-C_1L^{-1}R^{1/2}R_{ij}-
  C_2L^{-1}R^{1/2}(R_{ij}-{\textstyle{{1}\over{3}}}R\delta_{ij})
  \nonumber\\
  &&\qquad+C_3L^{-1}M^{1/2}M_{ij}-C_4L^{-1}R^{-1/2}MR_{ij},
\end{eqnarray}
\begin{eqnarray}
  \lefteqn{(\partial_t+u_k\partial_k)M_{ij}-
  M_{ik}\partial_ku_j-M_{jk}\partial_ku_i}&\nonumber\\
  &&=C_4L^{-1}R^{-1/2}MR_{ij}-(C_3+C_5)L^{-1}M^{1/2}M_{ij},
\end{eqnarray}
where $R=R_{ii}$ and $M=M_{ii}$ are the traces of the Reynolds and
Maxwell tensors, and $C_1$, \dots, $C_5$ are positive dimensionless
coefficients of a universal nature.

In the following subsections the physical origin and implications of
the various terms in the model will be made clear.

\subsection{Hydrodynamic decay and return to isotropy}

For freely decaying hydrodynamic turbulence in the absence of rotation
and a mean flow, one might start with a model of the form
\begin{equation}
  \partial_tR_{ij}=-C_1L^{-1}R^{1/2}R_{ij}.
\end{equation}
The turbulent kinetic energy density ${\textstyle{{1}\over{2}}}R$ then
decays monotonically and non-linearly according to
\begin{equation}
  \partial_tR=-C_1L^{-1}R^{3/2}.
\end{equation}
The functional form of the right-hand side is dictated by dimensional
considerations, as discussed above, and is also suggested by the form
of the triple correlations that appear in the exact equation
(\ref{rij_exact}).  The predicted decay of energy, $R\propto t^{-2}$
as $t\to\infty$, is faster than the behaviour $R\propto t^{-1}$
observed in turbulence generated by the passage of a stream of fluid
through a grid (Batchelor 1953).  The reason for this is that the
length-scale of the energy-containing eddies is constrained by the
constant $L$ in the present model, while it expands continuously
during the decay of grid-generated turbulence.

It is well known that freely decaying, anisotropic hydrodynamic
turbulence also exhibits a tendency to return to isotropy.  This
suggests the addition of a traceless term that redistributes energy
among the components of $R_{ij}$, i.e.
\begin{equation}
  \partial_tR_{ij}=-C_1L^{-1}R^{1/2}R_{ij}-
  C_2L^{-1}R^{1/2}(R_{ij}-{\textstyle{{1}\over{3}}}R\delta_{ij}).
\end{equation}
In this model, if initially $R>0$ but $R_{xx}=0$, for example, then
$R_{xx}$ will first grow and then decay.

\subsection{Turbulent Lorentz forces and small-scale dynamo action}

If the fluid is initially at rest and a random magnetic field is
introduced, the Lorentz forces will immediately drive turbulent
motions.  This is the origin of the term $C_3$, which appears as a
source for $R_{ij}$ and a sink for $M_{ij}$.  Note that a random
magnetic field that has a dominant $x$-component, say, drives motions
predominantly in the $x$-direction through the $\bB\cdot\nabla\bB$
force.\footnote{This may seem paradoxical as the total Lorentz force
  is orthogonal to $\bB$.  However, only the solenoidal part of the
  Lorentz force drives motion in an incompressible fluid, or in a
  compressible fluid under Boussinesq conditions.  The remaining
  gradient part is compensated by a pressure gradient.}  For this
reason the source for $R_{ij}$ has the tensorial form of $M_{ij}$.

The term $C_4$ is a source for $M_{ij}$ and a sink for $R_{ij}$.  It
can be considered as the effect of small-scale dynamo action.  Any
turbulent motion acts as a small-scale dynamo and causes a growth of
turbulent magnetic energy, at least when the field is weak.  This
occurs through the $\bB\cdot\nabla\bu$ term in the induction equation,
and therefore the source for $M_{ij}$ has the tensorial form of
$R_{ij}$.

\subsection{Energetics, equipartition and realizability}

The evolution of the turbulent kinetic and magnetic energy densities,
${\textstyle{{1}\over{2}}}R$ and ${\textstyle{{1}\over{2}}}M$, is
governed by the traces of the equations for $R_{ij}$ and $M_{ij}$.
Thus
\begin{eqnarray}
  \lefteqn{(\partial_t+u_i\partial_i)R=
  -2R_{ij}\partial_ju_i-C_1L^{-1}R^{3/2}+C_3L^{-1}M^{3/2}}&\nonumber\\
  &&-C_4L^{-1}R^{1/2}M,
  \label{ke}
\end{eqnarray}
\begin{eqnarray}
  \lefteqn{(\partial_t+u_i\partial_i)M=2M_{ij}\partial_ju_i+
  C_4L^{-1}R^{1/2}M}&\nonumber\\
  &&-(C_3+C_5)L^{-1}M^{3/2}.
  \label{me}
\end{eqnarray}
The total energy density satisfies
\begin{eqnarray}
  \lefteqn{(\partial_t+u_i\partial_i)
  ({\textstyle{{1}\over{2}}}R+{\textstyle{{1}\over{2}}}M)=
  (M_{ij}-R_{ij})\partial_ju_i}&\nonumber\\
  &&-C_1L^{-1}{\textstyle{{1}\over{2}}}R^{3/2}-
  C_5L^{-1}{\textstyle{{1}\over{2}}}M^{3/2}.
\end{eqnarray}
The first term on the right-hand side represents the extraction of
shear energy from the mean flow by the total turbulent stress
$M_{ij}-R_{ij}$.  The other two terms, $C_1$ and $C_5$, represent the
turbulent dissipation of kinetic and magnetic energy, and can be
thought of as operating through the formation of vortex sheets and
current sheets, respectively.

Terms $C_3$ and $C_4$ transfer energy between kinetic and magnetic
components.  The net rate of transfer from kinetic to magnetic energy
is
\begin{equation}
  (C_4R^{1/2}-C_3M^{1/2})L^{-1}{\textstyle{{1}\over{2}}}M
\end{equation}
and is positive or negative according to whether the kinetic or
magnetic energy dominates.  There is therefore a tendency towards
`equipartition' in the ratio $M/R=(C_4/C_3)^2$.

An important requirement of the model is that the kinetic and magnetic
energy densities should remain non-negative, not least because square
roots appear in the model.  In fact there is a stronger requirement,
mentioned above, because the Reynolds and Maxwell tensors are by
definition positive semi-definite tensors.  In the Appendix it is
shown that the positive semi-definite nature of $R_{ij}$ and $M_{ij}$
is preserved by the model, guaranteeing that they are realizable by
genuine velocity and magnetic fields.

\section{Non-linear outcome in the shearing sheet}

\label{Non-linear outcome in the shearing sheet}

In this section I examine the predictions of the model for the
incompressible shearing sheet.  The angular velocity and velocity
gradient are fixed by the parameters of the shearing sheet.  The
Reynolds and Maxwell tensors may be assumed to be independent of
position, corresponding to homogeneous turbulence.  The model then
reduces to a non-linear dynamical system describing the purely
temporal evolution of the stress tensors.  The non-trivial fixed
points of the dynamical system represent possible states of
statistically steady turbulence.

In numerical calculations I adopt the fiducial parameters
$C_1=C_2=C_3=C_4=C_5=1$, although other values have been experimented
with.  The fact that this simple choice gives reasonable results
indicates that no fine tuning is required for the model to behave
physically.  Ideally the parameters should be calibrated by comparison
with suitable numerical simulations.

\subsection{Freely decaying turbulence without rotation or shear}

With $\Omega=A=0$ there is nothing to initiate or sustain turbulence.
Any turbulence introduced into the system decays, tending to isotropy
as is does so.  For freely decaying isotropic turbulence
($R_{ij}={\textstyle{{1}\over{3}}}R\delta_{ij}$, etc.), the model
states that
\begin{eqnarray}
  \dot R&=&-C_1L^{-1}R^{3/2}+C_3L^{-1}M^{3/2}-C_4L^{-1}R^{1/2}M,\nonumber\\
  \dot M&=&C_4L^{-1}R^{1/2}M-(C_3+C_5)L^{-1}M^{3/2},
\end{eqnarray}
where the dot denotes differentiation with respect to time.  A typical
phase portrait is shown in Fig.~1.  All trajectories tend towards the
only fixed point, $R=M=0$, but do not reach it in a finite time.
During the decay there is a tendency towards equipartition.

\begin{figure}
  \centerline{\epsfbox{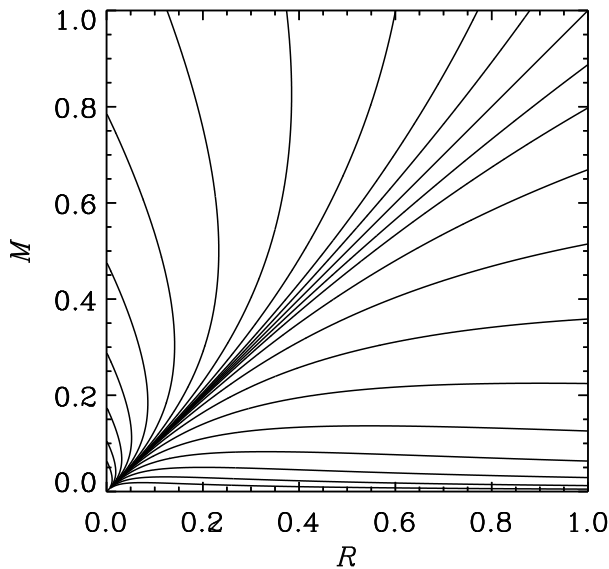}}
  \caption{Phase portrait for freely decaying isotropic turbulence in
    the absence of rotation or shear.  Fiducial parameters
    $C_1=C_3=C_4=C_5=1$ are adopted.  The diagram is self-similar and
    therefore the scale is arbitrary.  All trajectories tend towards
    the only fixed point, $R=M=0$.  During the decay there is a
    tendency towards equipartition.}
\end{figure}

\subsection{Non-rotating shear flow without a magnetic field}

The next simplest case to be examined is that of a non-rotating shear
flow (plane Couette flow, $\Omega=0$) with $M_{ij}=0$ throughout.
This case has been studied in the laboratory (with rigid boundaries)
and in numerical simulations with the periodic boundary conditions of
the shearing box (Pumir 1996).  At high Reynolds number, sustained
hydrodynamic turbulence develops.  The closure model ought to describe
this behaviour.

As described in Section~\ref{linear}, in the absence of rotation the
system is Rayleigh-neutral and the linear dynamics of $R_{ij}$ allows
for algebraic growth.  The outcome is determined by the non-linear
terms.  The system has only two fixed points, one of which is the
algebraically unstable trivial solution $R_{ij}=0$.  The non-trivial
fixed point represents a state of steady hydrodynamic turbulence and
is given by
\begin{eqnarray}
  &&R_{xx}=R_{zz}=\left({{C_2}\over{C_1+C_2}}\right){{1}\over{3}}R,
  \nonumber\\
  &&R_{yy}=\left({{3C_1+C_2}\over{C_1+C_2}}\right){{1}\over{3}}R,
  \nonumber\\
  &&R_{xy}={{C_1}\over{4LA}}R^{3/2},\nonumber\\
  &&R_{xz}=R_{yz}=0,
\end{eqnarray}
with
\begin{equation}
  R=\left[{{C_2}\over{C_1(C_1+C_2)^2}}\right]{{8}\over{3}}L^2A^2.
  \label{pcf}
\end{equation}
The existence of this non-trivial solution depends on both the shear
and the return-to-isotropy coefficient $C_2$.  The fixed point is
stable for all parameter values and is universally attracting.

A typical result of numerical integration of the time-dependent
equations is shown in Fig.~2.  The integration is started close to the
unstable fixed point $R_{ij}=0$ by introducing a small positive value
of $R_{xx}$ only.  Initially algebraic growth leads to a rapid
approach to the stable fixed point.

\begin{figure}
  \centerline{\epsfbox{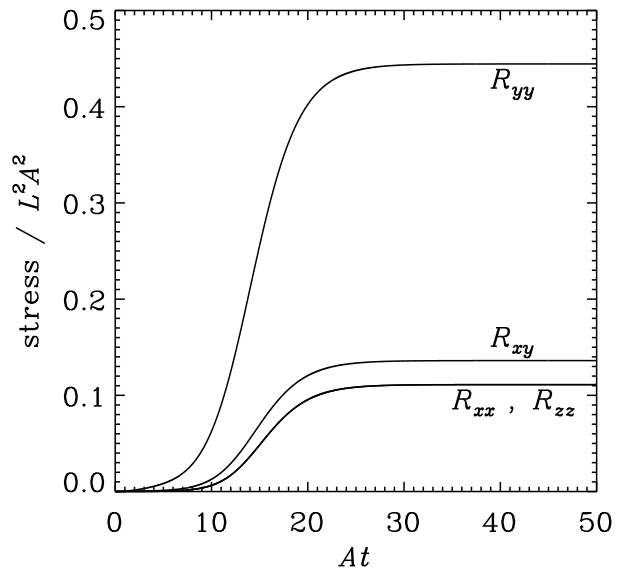}}
  \caption{Evolution of the Reynolds tensor in plane Couette flow,
    starting from a small hydrodynamic perturbation without a magnetic
    field.  Fiducial parameters $C_1=C_2=1$ are adopted.  The solution
    tends towards the stable fixed point representing a state of
    statistically steady hydrodynamic turbulence.  The components
    $R_{xz}$ and $R_{yz}$ are decoupled from the others, and tend to
    zero.}
\end{figure}

\subsection{Non-rotating shear flow with a magnetic field}

\label{Non-rotating shear flow with a magnetic field}

The dynamics changes when a magnetic field is introduced, because a
turbulent flow acts as a small-scale dynamo.  The non-trivial fixed
point representing a state of steady hydrodynamic turbulence is
unstable to a magnetic perturbation.  A new, stable fixed point
appears, corresponding to a state of steady MHD turbulence.

A typical result of numerical integration of the time-dependent
equations is shown in Fig.~3.  The integration is started close to the
trivial solution $R_{ij}=M_{ij}=0$ by introducing small positive
values of $R_{xx}$ and $M_{xx}$ only.  The system approaches the
stable fixed point representing magnetized turbulence.

\begin{figure*}
  \centerline{\epsfbox{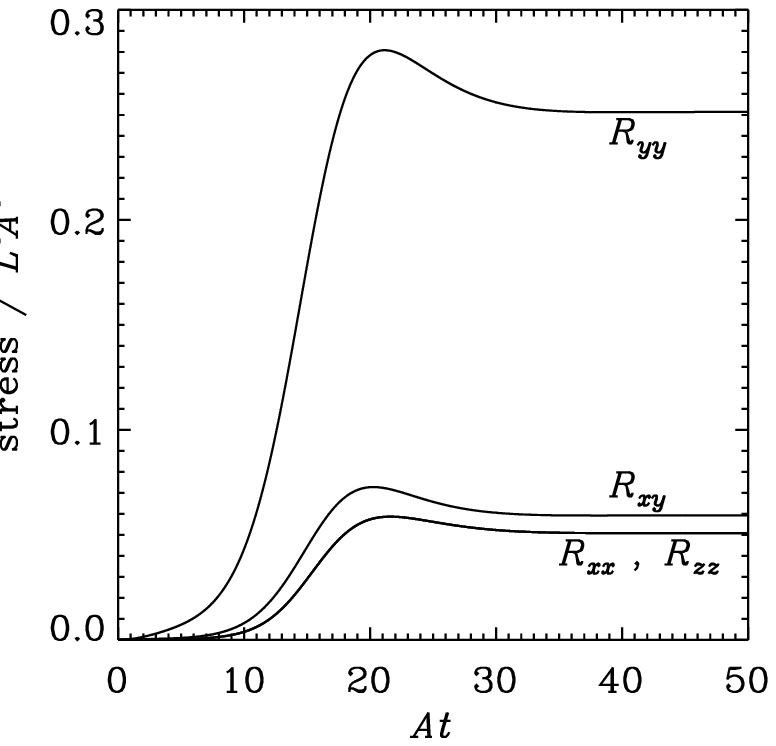}\qquad\epsfbox{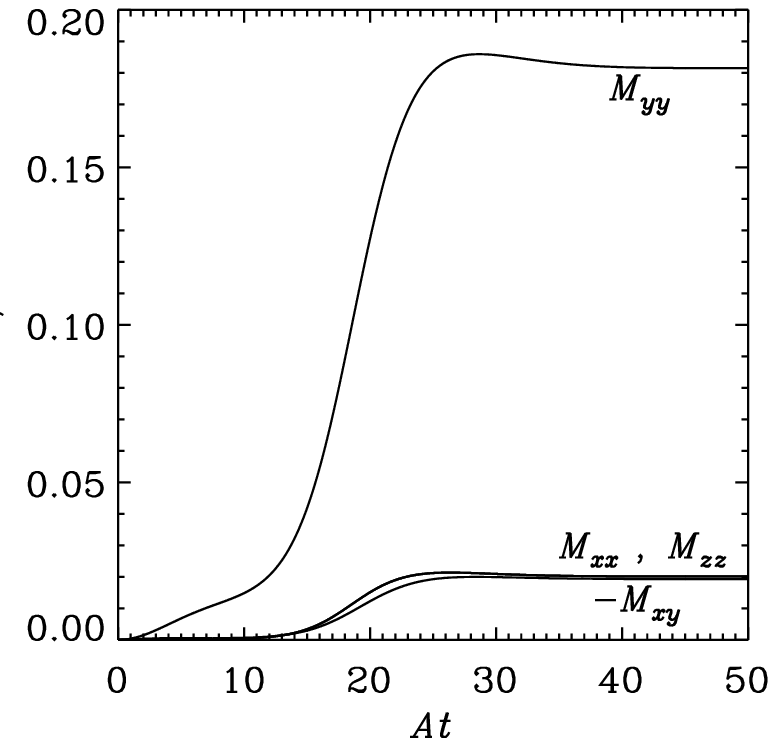}}
  \caption{Evolution of the Reynolds tensor (left) and Maxwell tensor
    (right) in plane Couette flow, starting from small hydrodynamic
    and magnetic perturbations.  Fiducial parameters
    $C_1=C_2=C_3=C_4=C_5=1$ are adopted.  The solution tends towards
    the stable fixed point representing a state of statistically
    steady MHD turbulence.  The solution approached in Fig.~2 is no
    longer attracting.  The components $R_{xz}$, $R_{yz}$, $M_{xz}$
    and $M_{yz}$ are decoupled from the others, and tend to zero.}
\end{figure*}

\subsection{Rotating shear flow without a magnetic field}

If one seeks a non-trivial fixed point representing a state of steady
hydrodynamic turbulence in rotating plane Couette flow, one obtains the
{\it formal\/} solution
\begin{eqnarray}
  &&R_{xx}=\left({{3\,{\rm Ro}^{-1}C_1+C_2}\over{C_1+C_2}}\right)
  {{1}\over{3}}R,
  \nonumber\\
  &&R_{yy}=\left[{{3(1-{\rm Ro}^{-1})C_1+C_2}\over{C_1+C_2}}\right]
  {{1}\over{3}}R,
  \nonumber\\
  &&R_{zz}=\left({{C_2}\over{C_1+C_2}}\right){{1}\over{3}}R,
  \nonumber\\
  &&R_{xy}={{C_1}\over{4LA}}R^{3/2},\nonumber\\
  &&R_{xz}=R_{yz}=0,
\end{eqnarray}
with
\begin{equation}
  R=\left[{{-6\,{\rm Ro}^{-1}({\rm Ro}^{-1}-1)C_1+C_2}
  \over{C_1(C_1+C_2)^2}}\right]{{8}\over{3}}L^2A^2,
\end{equation}
which is the generalization of equation (\ref{pcf}) for the
non-rotating case.  This solution is only meaningful, however, if
$R_{ij}$ is positive semi-definite, and this requires
\begin{equation}
  {{C_2}\over{C_1}}>6\,{\rm Ro}^{-1}({\rm Ro}^{-1}-1),
\end{equation}
as illustrated in Fig.~4.  (In the case of equality, the solution is
trivial.)  Where it exists, this solution appears to be stable.

\begin{figure}
  \centerline{\epsfbox{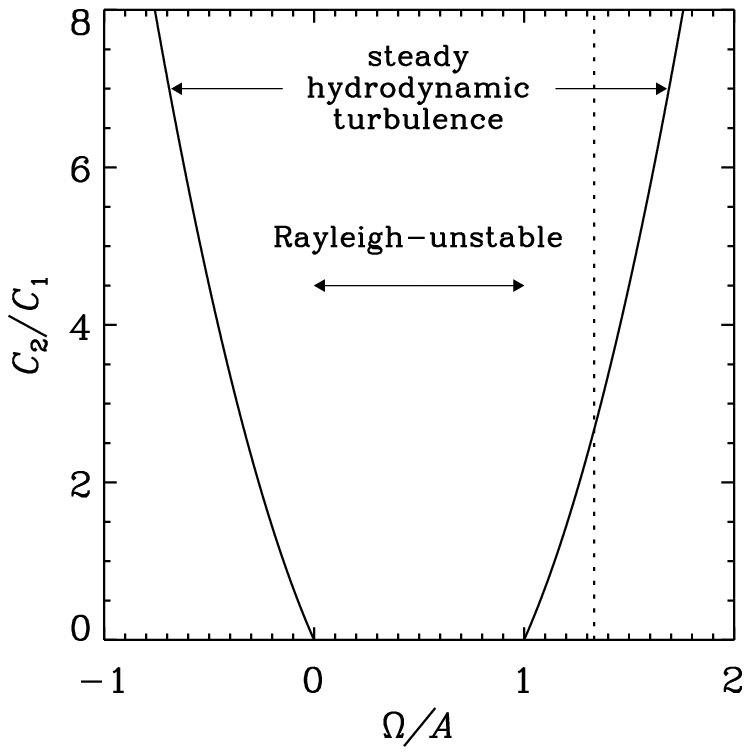}}
  \caption{Stability diagram for rotating plane Couette flow in the
    absence of a magnetic field.  The two parameters of the system are
    the inverse Rossby number $\Omega/A$ and the ratio $C_2/C_1$,
    which measures the tendency to return to isotropy.  When the
    system is Rayleigh-unstable ($0<\Omega/A<1$), or when the
    isotropizing tendency is sufficiently large, a stable state of
    steady hydrodynamic turbulence is permitted.  Otherwise the system
    exhibits decaying epicyclic oscillations when perturbed.  The
    Keplerian case $\Omega/A=4/3$ is shown with a dotted line.  Steady
    turbulence is permitted only if $C_2/C_1>8/3$.}
\end{figure}

For the Keplerian case ${\rm Ro}=3/4$ relevant to a standard accretion
disc, steady hydrodynamic turbulence is possible only when
$C_2/C_1>8/3$.  Although this is possible in principle, such a value
appears improbable a priori.  When $C_2/C_1=1$, for example, steady
turbulence is possible for $-0.145<{\rm Ro}^{-1}<1.145$.  For rotation
laws of the form $\Omega\propto r^{-q}$, this requires $q>1.746$.

Numerical integration of the time-dependent system confirms that,
when the system is perturbed from $R_{ij}=0$, it tends towards the
stable solution where this exists.  Otherwise it exhibits decaying
epicyclic oscillations about $R_{ij}=0$.

\subsection{Rotating shear flow with a magnetic field}

The introduction of a magnetic perturbation allows steady MHD
turbulence to develop in a rotating shear flow, even when Rayleigh's
stability criterion is amply satisfied, as in the Keplerian case ${\rm
  Ro}=3/4$ (Fig.~5).  This occurs through a non-linear
magnetorotational instability (non-linear because there is no imposed
magnetic flux) and the magnetic field is sustained through a
non-linear dynamo process.  The model therefore reproduces in broad
terms the finding of numerical studies of the non-linear evolution of
the magnetorotational instability (e.g. Hawley et al. 1995).

\begin{figure*}
  \centerline{\epsfbox{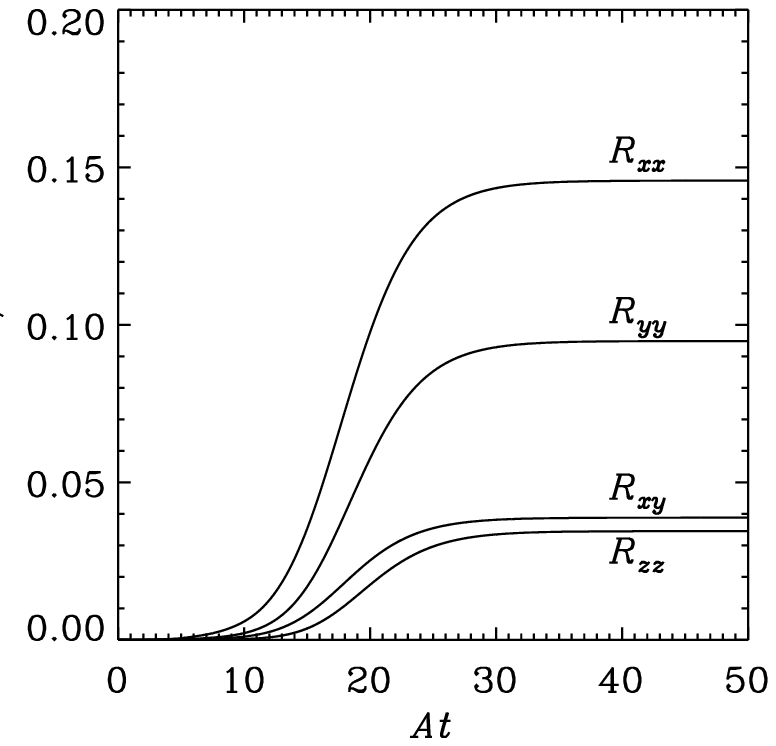}\qquad\epsfbox{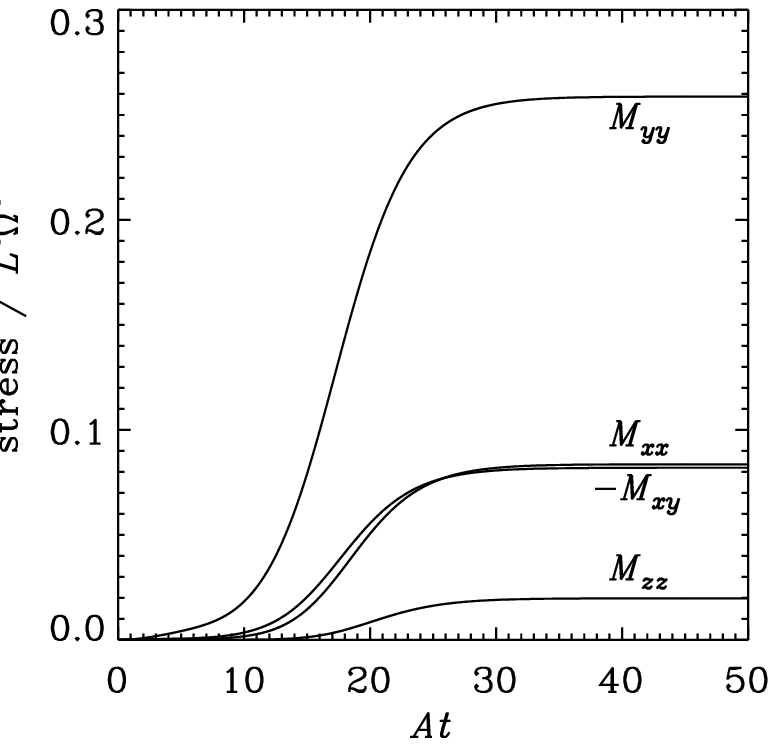}}
  \caption{Evolution of the Reynolds tensor (left) and Maxwell tensor
    (right) in a rotating shear flow with ${\rm Ro}=3/4$, starting
    from small hydrodynamic and magnetic perturbations.  Fiducial
    parameters $C_1=C_2=C_3=C_4=C_5=1$ are adopted.  The solution
    tends towards the stable fixed point representing a state of
    statistically steady MHD turbulence.  The components $R_{xz}$,
    $R_{yz}$, $M_{xz}$ and $M_{yz}$ are decoupled from the others, and
    tend to zero.}
\end{figure*}

\section{The governing equations in compressible flows}

\label{The governing equations in compressible flows}

The assumption of an incompressible fluid is useful in reducing the
problem to a minimal, although still formidable, complexity.  There
are two reasons why the model may not be suitable for immediate
application to a compressible fluid.  First, the mean density may be
non-uniform, especially if the mean velocity field has a non-zero
divergence.  Second, the turbulence may be essentially transonic,
changing the character of the motions and leading to greatly enhanced
dissipation through shock formation.  The second case is beyond the
scope of this paper and is unlikely to be important for the
magnetorotational instability, at least in the part of an accretion
disc where most of the mass resides.  The first case, however, is
important in a number of applications and can be treated in a simple
way.

In a compressible fluid it is more appropriate to define the mean
Reynolds and Maxwell tensors as
\begin{equation}
  R_{ij}=\langle\rho u_i'u_j'\rangle
\end{equation}
and
\begin{equation}
  M_{ij}=\Big\langle{{B_i'B_j'}\over{\mu_0}}\Big\rangle,
\end{equation}
which have the dimensions of stress and differ from the earlier
definitions by a factor of the density.  It is convenient not to
include the magnetic pressure perturbation in $M_{ij}$, but to regard
the mean magnetic stress as
$M_{ij}-{\textstyle{{1}\over{2}}}M\delta_{ij}$.

The first issue to consider is whether a divergence of the mean
velocity field, $\partial_iu_i$, should affect the evolution of
$R_{ij}$ or $M_{ij}$.  This velocity divergence does not appear in the
equation of motion, but does appear in the equation of mass
conservation and the induction equation in the forms
$\partial_t\rho=\cdots-\rho\partial_iu_i$ and
$\partial_tB_i=\cdots-B_i\partial_ju_j$ respectively.  This motivates
the addition of terms in the forms
$\partial_tR_{ij}=\cdots-R_{ij}\partial_ku_k$ and
$\partial_tM_{ij}=\cdots-2M_{ij}\partial_ku_k$ respectively.

The second issue concerns the definition of the vertical length-scale
$L$ in a stratified disc, and the effect of the stratification on the
vertical profile of the stress.  In the incompressible system the
turbulence is homogeneous and the stress independent of $z$.
Numerical simulations of stratified, isothermal accretion discs
(Miller \& Stone 2000) suggest that the stress is also stratified,
being roughly proportional to the density (or pressure), and in many
applications it is analytically convenient to assume that the stress
scales with the pressure.  This can be achieved in a natural way by
identifying the vertical length-scale as $L=c_{\rm s}/\Omega_z$, where
$c_{\rm s}=(p/\rho)^{1/2}$ is the isothermal sound speed and
$\Omega_z$ the vertical oscillation frequency, which measures the
curvature of the gravitational potential and is equal to $\Omega$ in a
Keplerian disc.  In an isothermal disc $L$ is then equal to the
Gaussian scale-height $H$.

Accordingly, the system of equations recommended for a compressible
accretion flow is as follows, written in an inertial frame of
reference.  The equation of mass conservation,
\begin{equation}
  \partial_t\rho+\partial_i(\rho u_i)=0.
  \label{compressible_rho}
\end{equation}
The equation of motion,
\begin{equation}
  \rho(\partial_t+u_j\partial_j)u_i=-\rho\partial_i\Phi-
  \partial_i(p+{\textstyle{{1}\over{2}}}M)+\partial_j(M_{ij}-R_{ij}).
\end{equation}
The equation for the Reynolds tensor,
\begin{eqnarray}
  \lefteqn{(\partial_t+u_k\partial_k)R_{ij}+R_{ik}\partial_ku_j+
  R_{jk}\partial_ku_i+R_{ij}\partial_ku_k=}&\nonumber\\
  &&\Omega_zp^{-1/2}\left[-C_1R^{1/2}R_{ij}-
  C_2R^{1/2}(R_{ij}-{\textstyle{{1}\over{3}}}R\delta_{ij})\right.\nonumber\\
  &&\left.\qquad+C_3M^{1/2}M_{ij}-C_4R^{-1/2}MR_{ij}\right].
\end{eqnarray}
The equation for the Maxwell tensor,
\begin{eqnarray}
  \lefteqn{(\partial_t+u_k\partial_k)M_{ij}-M_{ik}\partial_ku_j-
  M_{jk}\partial_ku_i+2M_{ij}\partial_ku_k=}&\nonumber\\
  &&\Omega_zp^{-1/2}\left[C_4R^{-1/2}MR_{ij}-(C_3+C_5)M^{1/2}M_{ij}\right].
  \label{compressible_mij}
\end{eqnarray}
The thermal energy equation,
\begin{equation}
  \rho T(\partial_t+u_i\partial_i)s=\Omega_zp^{-1/2}
  \left({\textstyle{{1}\over{2}}}C_1R^{3/2}+
  {\textstyle{{1}\over{2}}}C_5M^{3/2}\right)-\partial_iF_i.
  \label{compressible_s}
\end{equation}
Here $T$ is the temperature, $s$ the specific entropy and $F_i$ the
radiative energy flux.  The positive contributions on the right-hand
side represent turbulent heating through viscous and resistive decay.

The total energy is then exactly conserved in the form
\begin{eqnarray}
  \lefteqn{\partial_t\left[\rho({\textstyle{{1}\over{2}}}u^2+\Phi+e)+
  {\textstyle{{1}\over{2}}}R+{\textstyle{{1}\over{2}}}M\right]}&\nonumber\\
  &&+\partial_i\left[\rho({\textstyle{{1}\over{2}}}u^2+\Phi+e)u_i+pu_i+
  ({\textstyle{{1}\over{2}}}R+M)u_i\right.\nonumber\\
  &&\left.\qquad+(R_{ij}-M_{ij})u_j+F_i\right]=0,
\end{eqnarray}
provided that the gravitational potential $\Phi$ is independent of
$t$.  Here $e$ is the specific internal energy, such that ${\rm
  d}e=T\,{\rm d}s-p\,{\rm d}(\rho^{-1})$.  The existence of this
conservation law, with positive-definite turbulent energy densities
and heating rates, implies a certain self-consistency in the equations
of the model.  The terms that were added in passing to the
compressible model are required to have the form that they do in order
that energy be conserved.

It is worth remarking that when $C_1=\cdots=C_5=0$ and $R_{ij}=0$,
these equations reduce exactly to those of ideal MHD if one identifies
$M_{ij}$ as the Maxwell tensor $B_iB_j/\mu_0$ of an arbitrary
large-scale magnetic field $B_i$ advected by the mean flow $u_i$.  For
the magnetic terms in the equation of motion are precisely the Lorentz
force of such a magnetic field, and the equation for the Maxwell
tensor is equivalent to the induction equation of ideal MHD.
Furthermore, the magnetic energy density,
${\textstyle{{1}\over{2}}}M=B^2/2\mu_0$, and the Poynting flux,
$Mu_i-M_{ij}u_j=(B^2u_i-B_iB_ju_j)/\mu_0=(\bE\times\bB)_i/\mu_0$, can
both be identified in the energy equation.

\section{Applications and implications}

\label{Applications and implications}

\subsection{Character of the equations}

The governing equations set out in the previous section usually have
the formal character of a hyperbolic system, indicating that
information is propagated in a causal manner at finite speeds.  To see
this, consider the behaviour of infinitesimal wavelike perturbations
of an arbitrary solution of the equations.  Let the perturbations have
the form of a rapidly varying phase factor, $\exp[{\rm
  i}\varphi(\bx,t)]$, multiplied by slowly varying functions of $\bx$
and $t$.  The wavenumber $\bk=\nabla\varphi$ and frequency
$\omega=-\partial_t\varphi$ are assumed to be such that $|\bk|^{-1}$
and $|\omega|^{-1}$ are small compared to the typical length-scales
and time-scales of the basic state.  Although the model is not
necessarily valid in this regime, such an analysis serves to uncover
the formal mathematical structure of the equations.

By linearizing equations
(\ref{compressible_rho})--(\ref{compressible_s}) and eliminating
perturbations in favour the velocity perturbation $u_i'$, one obtains
the algebraic eigenvalue problem
\begin{eqnarray}
  \lefteqn{\rho(\omega-k_ju_j)^2u_i'=(R_{jk}+M_{jk})k_jk_ku_i'}&\nonumber\\
  &&+(\rho v_{\rm s}^2k_ik_j+2R_{ij}k_jk_k)u_j'\nonumber\\
  &&+(Mk_ik_j-M_{ik}k_jk_k-M_{jk}k_ik_k)u_j'
  \label{dispersion}
\end{eqnarray}
for the local WKB modes of the system, where $v_{\rm s}^2=(\partial
p/\partial\rho)_s$ is the square of the adiabatic sound speed.  As in
compressible MHD, the WKB dispersion relation has 6 frequency
eigenvalues $\omega$ for any choice of the real wavevector $\bk$.  If
any of the eigenvalues has a non-zero imaginary part, the system
experiences a local instability.  Otherwise the eigenvalues are all
real and the group velocities $\partial\omega/\partial k_i$ are
independent of $|\bk|$, indicating that the wave propagation in this
limit is anisotropic but non-dispersive.  The system of equations is
then formally hyperbolic.

A full investigation of the dispersion relation is difficult but the
following observations may be made.  First, when $R_{ij}=M_{ij}=0$,
the dispersion relation is the standard one of compressible
hydrodynamics, and information is propagated at the sound speed
relative to the mean flow.  Second, when $R_{ij}=0$ but $M_{ij}\ne0$,
the system can be shown to be hyperbolic.  The dispersion relation is
related to that of compressible MHD.  Third, in an incompressible
fluid, modes with $k_iu_i'\ne0$ are eliminated as their speed of
propagation is infinite.  The remaining part of the dispersion
relation, given by the first line of equation (\ref{dispersion}),
describes the propagation of transverse modes, similar to Alfv\'en
waves, at finite speed.  Finally, there are circumstances in which the
system fails to be hyperbolic because the dispersion relation
indicates a local instability, but this appears to be atypical.

\subsection{Stratified shearing sheet}

In a stratified shearing sheet, magnetorotational turbulence would
develop according to these equations just as in an incompressible
shearing sheet, except that all stress components would be
proportional to $p/\Omega_z^2$ rather than $L^2$.  The total shear
stress $M_{xy}-R_{xy}$ can be compared with that corresponding to an
effective viscosity $\mu=\alpha p/\Omega$.  For the fiducial
parameters in a Keplerian disc, this gives $\alpha\approx0.081$.  If
all the parameters $C_i$ are scaled by a constant factor, this value
of $\alpha$ scales as $1/C_i^2$.

\subsection{Non-Keplerian rotation}

When the Rossby number ${\rm Ro}=A/\Omega$ of the shearing sheet is
varied in the range $0<{\rm Ro}<1$ in which the system is
Rayleigh-stable but magnetorotationally unstable, the total shear
stress $M_{xy}-R_{xy}$ in the steady turbulent state is not simply
proportional to the shear rate $2A$, at fixed angular velocity.  This
is shown in Fig.~6, and demonstrates one aspect in which the present
model differs significantly from a viscous representation of the
turbulence.

\begin{figure}
  \centerline{\epsfbox{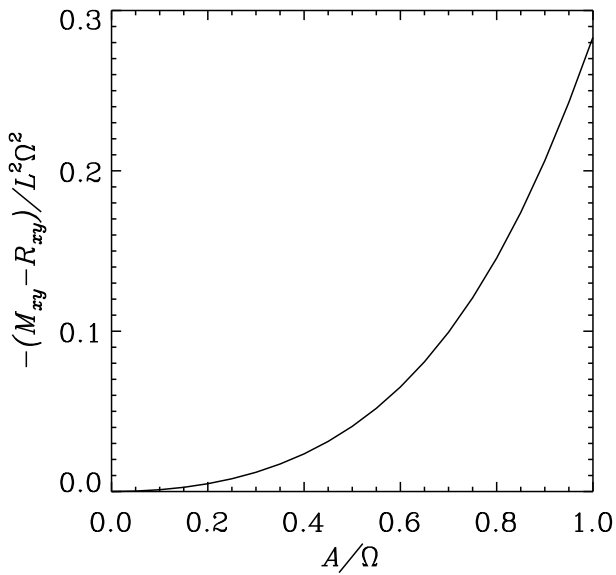}}
  \caption{The non-linear relation between the total shear stress in
    steady magnetorotational turbulence and the shear rate, at fixed
    angular velocity.  Fiducial parameters $C_1=C_2=C_3=C_4=C_5=1$ are
    adopted.}
\end{figure}

\subsection{Damping of the two-dimensional wave}

\label{Damping of the two-dimensional wave}

The local dispersion relation for axisymmetric density waves in a
two-dimensional disc model is
\begin{equation}
  \omega^2=\kappa^2+v_{\rm s}^2k^2,
  \label{gt}
\end{equation}
where $k$ is the radial wavenumber (Goldreich \& Tremaine 1979).  The
same mode can be identified within the present model if one considers
a three-dimensional compressible shearing sheet without vertical
gravity and stratification.  To avoid complications it is convenient
to replace the thermal energy equation by the isothermal condition
$p=c_{\rm s}^2\rho$, $c_{\rm s}={\rm constant}$.  Equations
(\ref{compressible_rho})--(\ref{compressible_mij}), linearized around
a basic state consisting of steady, homogeneous magnetorotational
turbulence, then admit solutions proportional to $\exp({\rm i}kx-{\rm
  i}\omega t)$, with no perturbations of
$(u_z,R_{xz},R_{yz},M_{xz},M_{yz})$.  In the absence of turbulent
stresses, the dispersion relation (\ref{gt}) is recovered.  A
numerical solution of the dispersion relation in the presence of
turbulent stresses indicates that the two-dimensional wave is damped
in a Keplerian disc when the fiducial parameters $C_i=1$ are adopted
(Fig.~7).  The real part of the frequency is somewhat greater than
predicted by equation (\ref{gt}), and agrees much better if $v_{\rm
  s}$ is replaced by the `magnetosonic speed' $(c_{\rm
  s}+M/\rho)^{1/2}$.  Interestingly, the imaginary part corresponds
quite closely with a viscous damping rate $\nu k^2$ if the effective
viscosity is $\nu=\alpha c_{\rm s}^2/\Omega$ with $\alpha\approx0.06$.
This is remarkable when it is considered that there are no viscous or
diffusive terms in the governing equations.

\begin{figure*}
  \centerline{\epsfbox{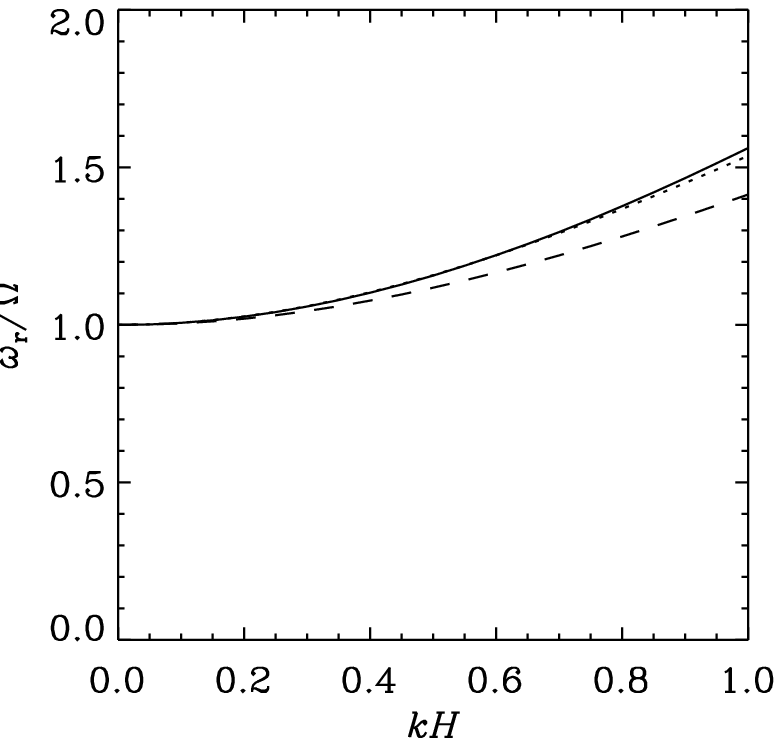}\qquad\epsfbox{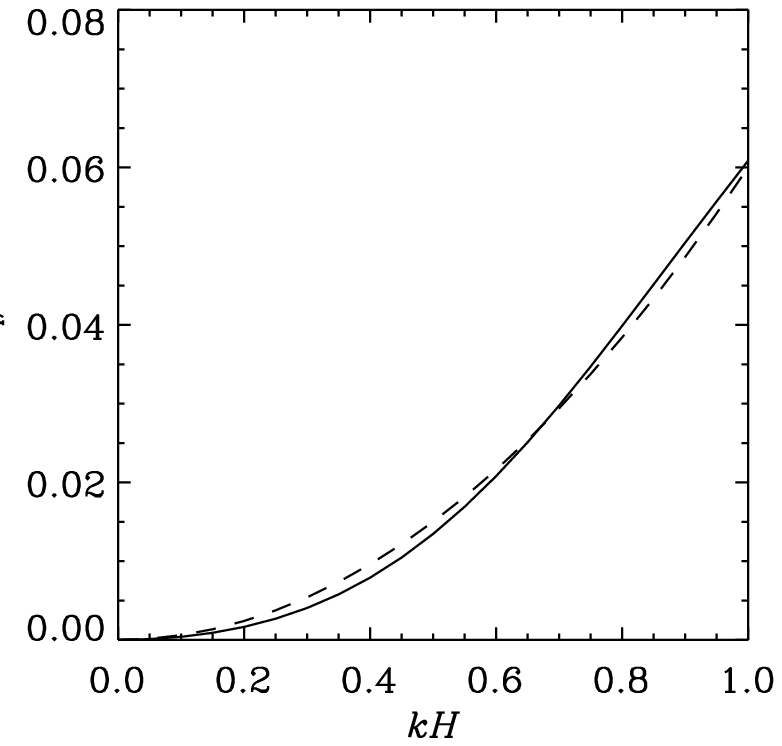}}
  \caption{Dispersion relation of the two-dimensional wave in a
    three-dimensional, compressible shearing sheet without
    stratification.  Keplerian rotation and isothermal perturbations
    are assumed, and fiducial parameters $C_1=C_2=C_3=C_4=C_5=1$ are
    adopted.  Left: real part of the frequency (solid line), compared
    with the standard dispersion relation (\ref{gt}) (dashed line).
    Much better agreement is found when the magnetosonic speed is used
    in place of the sound speed (dotted line).  Right: damping rate of
    the wave (solid line), compared with the expectation based on a
    viscous damping corresponding to $\alpha=0.06$ (dashed line).}
\end{figure*}

\subsection{Damping of shearing epicyclic motions}

\label{Damping of shearing epicyclic motions}

When a stratified disc is warped, horizontal pressure gradients are
introduced that drive epicyclic motions with an amplitude proportional
to the distance $z$ above the mid-plane (Papaloizou \& Pringle 1983).
The amplitude and phase of these oscillations are critical to the
propagation and damping of the warp.  As discussed by Torkelsson et
al. (2000), the dynamics of these shearing epicyclic motions, and in
particular their damping time-scale, can be studied within the
shearing sheet.

Consider equations (\ref{compressible_rho})--(\ref{compressible_mij}),
linearized around a basic state consisting of steady, homogeneous
magnetorotational turbulence, in which the perturbations depend only
on $z$ and $t$.  It is possible to arrange for all perturbations to
vanish other than those of $(u_x,u_y,R_{xz},R_{yz},M_{xz},M_{yz})$.
Then
\begin{equation}
  \rho(\partial_tu_x'-2\Omega u_y')=\partial_z(M_{xz}'-R_{xz}'),
\end{equation}
\begin{equation}
  \rho\left[\partial_tu_y'+2(\Omega-A)u_x'\right]=\partial_z(M_{yz}'-R_{yz}'),
\end{equation}
\begin{eqnarray}
  \lefteqn{\partial_tR_{xz}'-2\Omega R_{yz}'+R_{zz}\partial_zu_x'=}
  &\nonumber\\
  &&\Omega_zp^{-1/2}\left[-(C_1+C_2)R^{1/2}R_{xz}'+C_3M^{1/2}M_{xz}'
  \right.\nonumber\\
  &&\left.-C_4R^{-1/2}MR_{xz}'\right],
\end{eqnarray}
\begin{eqnarray}
  \lefteqn{\partial_tR_{yz}'+2(\Omega-A)R_{xz}'+R_{zz}\partial_zu_y'=}
  &\nonumber\\
  &&\Omega_zp^{-1/2}\left[-(C_1+C_2)R^{1/2}R_{yz}'+C_3M^{1/2}M_{yz}'
  \right.\nonumber\\
  &&\left.-C_4R^{-1/2}MR_{yz}'\right],
\end{eqnarray}
\begin{eqnarray}
  \lefteqn{\partial_tM_{xz}'-M_{zz}\partial_zu_x'=}
  &\nonumber\\
  &&\Omega_zp^{-1/2}\left[C_4R^{-1/2}MR_{xz}'-(C_3+C_5)M^{1/2}M_{xz}'\right],
\end{eqnarray}
\begin{eqnarray}
  \lefteqn{\partial_tM_{yz}'+2AM_{xz}'-M_{zz}\partial_zu_y'=}
  &\nonumber\\
  &&\Omega_zp^{-1/2}\left[C_4R^{-1/2}MR_{yz}'-(C_3+C_5)M^{1/2}M_{yz}'\right].
\end{eqnarray}
These equations admit solutions in which all perturbations are
proportional to $\exp(-{\rm i}\omega t)$, while $u_x'$ and $u_y'$ are
proportional to $z$ and $R_{xz}'$, $R_{yz}'$ $M_{xz}'$ and $M_{yz}'$
are proportional to $p$.  The six eigenvalues $\omega$, corresponding
to three physically distinct modes, follow from an algebraic
eigenvalue problem that is easily solved numerically.  Two of these
modes are strongly damped, while the other represents a slowly damped
shearing epicyclic motion.  For a Keplerian disc and fiducial
parameters $C_i=1$, the damping rate is $0.0254\Omega$, the same that
would be obtained from an alpha viscosity if $\alpha=0.0254$.  This is
smaller than the value of $0.081$ required to account for the shear
stress $M_{xy}-R_{xy}$.

\section{Comparison with previous work}

\label{Comparison with previous work}

\subsection{Comparison with analytical models by other authors}

The historical development of closed model equations for the stress
tensor in a turbulent fluid is described in the review article of
Speziale (1991).  Early studies were based on the concepts of eddy
viscosity and mixing length introduced by Boussinesq and Prandtl.  A
systematic investigation of the equation for the mean Reynolds stress
was initiated in the 1950s.  Out of this arose (among others) the
$K$--$\epsilon$ model, which is widely used in engineering
applications despite its numerous deficiencies.  More recent
approaches have aimed at a greater fidelity to experimental and
numerical results, typically by elaborating algebraic models of the
pressure--strain correlation and trying to fix the coefficients
therein.

The present work is related to these Reynolds-stress closure models in
that I have worked with the exact equations for the mean Reynolds and
Maxwell tensors, retaining the form of the linear terms and proposing
closures for the non-linear terms.  However, I have tried to take a
fundamentally different approach to the non-linear effects by
recognizing the essential role of the length-scale $L$ in the
saturation of the turbulence and by identifying each non-linear term
with a known physical process.  In the absence of a magnetic stress,
the equations adopted here for the Reynolds tensor are considerably
simpler than those advocated by, e.g., Speziale (1991), yet they do
ensure realizability, allow for the return to isotropy, and may well
give a more accurate representation of the linear and non-linear
stability properties of rotating shear flows.

The application of Reynolds-stress closure models to accretion discs
has been proposed in a series of papers by Kato (1994), Kato \&
Inagaki (1994) and Kato \& Yoshizawa (1993, 1995, 1997).  In some of
these papers the Maxwell stress is also modelled.  While these studies
have a fair amount in common with the present approach, some important
differences must be emphasized.  In the absence of a magnetic stress,
Kato \& Yoshizawa (1997) have argued that steady hydrodynamic
turbulence may be sustained in a Keplerian accretion disc.  In fact,
their closure model is based on one by Launder, Reece \& Rodi (1975),
which predicts stability for a Keplerian shear flow (see Speziale
1991); only by modifying one of the terms were Kato \& Yoshizawa able
to find a steady turbulent state.  More importantly, the work of Kato
and co-workers does not address the issue of the non-linear saturation
of hydrodynamic or MHD turbulence.  Although it predicts the relative
magnitudes of the various components of the Reynolds (and Maxwell)
tensors, it does not predict the magnitude of the turbulent energy in
the saturated state.

Some further connections with the astrophysical literature can be
made.  Schramkowski et al. (1996) devised a kinematic mean-field
theory for the evolution of the Maxwell tensor in a turbulent
accretion disc, as an alternative to studying the mean magnetic field
itself.  Their theory is covariant and includes the interaction of the
Maxwell tensor with the mean velocity field.  In addition to the
`alpha' and `beta' effects of standard mean-field electrodynamics,
they emphasized the role of the `gamma' effect by which a turbulent
flow amplifies small-scale magnetic energy.  This can be related to
the $C_4$ term of the present model.

Soon after Balbus \& Hawley (1991) first wrote on the
magnetorotational instability in accretion discs, Tout \& Pringle
(1992) presented a simple model of a magnetic dynamo cycle that uses
the differential rotation to generate azimuthal field from radial
field, the Parker instability to convert azimuthal field into vertical
field, and the magnetorotational instability to regenerate radial
field from vertical field, while magnetic reconnection provides
dissipation.  In effect, they wrote down a non-linear dynamical system
for the three field components, which could be thought of as either
mean or RMS values.  The present model is ultimately based on a
similar idea, although it treats the Maxwell stress covariantly as a
single tensor object.  The Parker instability (magnetic buoyancy) does
not play a role here, and the magnetorotational instability is an
outcome of the model rather than an ingredient.

\subsection{Comparison with a viscoelastic model}

In recent work on the dynamics of eccentric discs I questioned the use
of a viscous model of the turbulent stress in an accretion disc
(Ogilvie 2001).  The assumption that the stress is linearly and
instantaneously related to the rate of strain typically leads to an
instability whereby an initially circular disc develops eccentricity
on short radial length-scales (see also Lyubarskij, Postnov \&
Prokhorov 1994); in fact, this is essentially the same effect as the
viscous overstability of axisymmetric waves discovered by Kato (1978).
In order to take account of the non-zero relaxation time $\tau$ of the
turbulence, I proposed the use of a viscoelastic model for the Maxwell
stress tensor,
\begin{equation}
  M_{ij}+\tau\sD M_{ij}=\mu(\partial_iu_j+\partial_ju_i)+
  \left(\mu_{\rm b}-{\textstyle{{2}\over{3}}}\mu\right)
  (\partial_ku_k)\delta_{ij},
\end{equation}
where $\mu$ is the effective shear viscosity, $\mu_{\rm b}$ the
effective bulk viscosity, and $\sD$ is a time-derivative operator such
that $\sD M_{ij}$ is equal to the left-hand side of equation
(\ref{compressible_mij}).

This model is a compressible version of the upper-convected Maxwell
fluid of non-Newtonian fluid dynamics, and is motivated by the fact
that the Maxwell tensor satisfies the equation $\sD M_{ij}=0$ exactly
in ideal MHD.  The analogy between viscoelasticity and MHD has been
developed in some detail by Ogilvie \& Proctor (2002), who show that
the magnetorotational instability is also found in viscoelastic
Couette flow.  The introduction of the relaxation term, with $\tau$
being comparable to the dynamical time-scale $\Omega^{-1}$, has a
profound influence on the dynamics of eccentric discs and tends to
remove the viscous overstability.

The viscoelastic model was intended to introduce the effect of the
relaxation time in the simplest possible way, and generalizes the
conventional viscous model by the introduction of a single new
parameter, the Weissenberg number $\Omega\tau$.  It is not a true
magnetorotational model because it neglects the Reynolds stress and
relies on an effective viscosity to generate a non-zero Maxwell
stress.  In contrast, the model proposed in the present paper does not
require any artificial source terms because the stresses arise through
an instability of the trivial state and through a cooperative
non-linear interaction between the Reynolds and Maxwell stresses.  The
cost paid for this is the need for approximately twice as many terms,
variables and parameters in the model.  Nevertheless, the viscoelastic
model can be compared closely with equation (\ref{compressible_mij})
if one identifies $\tau^{-1}=(C_3+C_5)\Omega_z(M/p)^{1/2}$ and
understands that the $C_4$ term, in acting as a source for $M_{ij}$,
is analogous to the effective viscosity.

\subsection{Comparison with numerical simulations}

The behaviour of the present model in the shearing sheet, as set out
in the various parts of Section~\ref{Non-linear outcome in the
  shearing sheet}, agrees at least qualitatively with the results of
existing numerical simulations.  For a non-rotating shear flow without
a magnetic field, one may compare with the simulations by Pumir
(1996), which demonstrate the development of statistically steady,
homogeneous and anisotropic turbulence.  A detailed comparison with
the mean stresses obtained in numerical simulations, preferably in the
limit of horizontally extended shearing boxes $L_x, L_y\gg L_z$, would
test the accuracy of the model and constrain the values of the
parameters.

Rotating shear flows without a magnetic field have been studied
numerically by Hawley, Balbus \& Winters (1999).  They investigated
rotation laws of the form $\Omega\propto r^{-q}$, finding that
Keplerian shear flows ($q=3/2$) are hydrodynamically stable while
constant-angular momentum shear flows ($q=2$) are non-linearly unstable
and become turbulent.  They tried to locate the critical value of $q$
at which hydrodynamic turbulence could just be sustained, and found it
to be close to $1.95$, although this may depend to some extent on the
resolution of the numerical method and the initial conditions adopted.
These results are in qualitative agreement with Fig.~4, which also
suggests that hydrodynamic turbulence can be sustained only in flows
that are Rayleigh-unstable or only just Rayleigh-stable.

The present model predicts the development of steady MHD turbulence in
a Keplerian shear flow, much as seen in numerical simulations by
Hawley et al. (1995), although it is more directly applicable to
calculations without an imposed magnetic flux, such as those of
Brandenburg et al. (1995) and Stone et al. (1996).  The initial
magnetic field in the latter simulations is well ordered, although of
zero mean, and gives rise to an initial phase of exponential growth
that is not captured in the model.  Nevertheless, the properties of
the saturated stresses appear to be in good qualitative agreement.  A
detailed numerical comparison would be inappropriate, because
unfortunately there is no consensus as to the `correct' absolute or
relative magnitudes of the mean stress components in a Keplerian shear
flow.  For example, Brandenburg et al. (1995) and Stone et al. (1996),
although adopting identical initial conditions and numerical
resolutions, obtain widely differing quantitative results; in
particular, they disagree on whether the turbulent kinetic energy or
the turbulent magnetic energy is greater.  It is unclear whether this
discrepancy is attributable to the difference in boundary conditions
or to a lack of numerical convergence.

Abramowicz et al. (1996) have simulated the magnetorotational
instability in non-Keplerian shear flows, having in mind the
application to the inner parts of accretion discs around black holes.
They found a non-linear relation between the shear stress and the
shear rate, at fixed angular velocity, in good agreement with Fig.~6.

Interestingly, the present model also allows for sustained MHD
turbulence in the case of negative Rossby number, $A/\Omega<0$, which
corresponds to a situation in which the angular velocity increases
outwards.  This regime may appear astrophysically improbable but may
be relevant to the boundary layer between a star and a disc.  While
boundary layers have been simulated in recent global studies (Armitage
2002; Steinacker \& Papaloizou 2002), the case of negative Rossby
number does not appear to have been explored in shearing-box
simulations.  There is no linear magnetorotational instability in this
case (except when the Hall effect is important, cf.  Balbus \& Terquem
2001), but the present model suggests that a turbulent MHD state may
nevertheless exist and is accessed by a non-linear instability of the
basic state.  The energetic arguments presented by Balbus \& Hawley
(1998) appear to allow for this possibility, and it would be valuable
to test this hypothesis in numerical simulations.

In one study, Hawley, Gammie \& Balbus (1996) repeated one of their
typical shearing-box simulation of magnetorotational turbulence, but
omitted the Coriolis force, thereby converting the calculation into a
local model of a non-rotating shear flow.  They found that the
magnetic energy was not sustained, although the turbulent kinetic
energy continued to grow until the calculation was halted.  This
differs from the prediction of the present model in
Section~\ref{Non-rotating shear flow with a magnetic field}, that
turbulent plane Couette flow acts as a small-scale dynamo and allows
the magnetic energy to grow to a substantial fraction of equipartition
with the turbulent kinetic energy, which itself reaches a saturated
value.  Indeed, the model is constructed on the basis that any
three-dimensional turbulent flow at sufficiently high ${\rm Rm}$, by
virtue of its stretching property, tends to amplify a weak,
small-scale magnetic field introduced into it, at least until the
Lorentz forces can modify the flow.  In plane Couette flow, the
shearing background provides a further means of amplifying the
magnetic energy.  The numerical result of Hawley et al. (1996), that
the combination of turbulence and shear has the opposite
characteristic, is therefore surprising.  One possible explanation for
the discrepancy between the numerical result and the prediction of the
model is that ${\rm Rm}$ is too small in the simulation to capture the
small-scale dynamo.  Alternatively, it may indicate a deficiency in
the model.  Further numerical studies may resolve this question.

Regrettably, there are no published numerical studies of the
propagation and damping of the two-dimensional wave in turbulent
discs, to compare with the predictions of Section~\ref{Damping of the
  two-dimensional wave}.  This problem is of some importance,
especially in view of its application to tidally generated waves in
binary stars and protoplanetary systems, and would make a further test
of the present model.

However, Torkelsson et al. (2000) have already studied numerically the
interaction of magnetorotational turbulence with the shearing
epicyclic motions found in warped discs (cf.~Section~\ref{Damping of
  shearing epicyclic motions}).  They found that the epicyclic motions
are damped, but at a rate approximately one-half that expected from
the action of an isotropic viscosity (the viscosity being measured
from the total $xy$-stress).  This is similar to the predictions of
the present model as set out in Section~\ref{Damping of shearing
  epicyclic motions}.

\section{Conclusion}

\label{Conclusion}

The representation of the turbulent stresses in an accretion disc by
the classical alpha viscosity model of Shakura \& Sunyaev (1973) has
been extremely successful in allowing the theory of accretion discs to
develop even in the absence of a proper understanding of the
turbulence, and in making possible a quantitative comparison between
theory and observations.  However, the magnetorotational instability
is now widely accepted to be the origin of the turbulence in
sufficiently ionized discs (e.g. Balbus \& Hawley 1998), and the alpha
viscosity model fails to describe numerous aspects of this process.
In this paper I have introduced a new analytical model that aims to
represent more faithfully the dynamics of magnetorotational turbulent
stresses and bridge the gap between analytical studies and numerical
simulations.  Covariant evolutionary equations for the mean Reynolds
and Maxwell tensors are presented, which correctly include the linear
interaction with the mean flow.  Non-linear and dissipative effects,
in the absence of an imposed magnetic flux and in the limit of large
Reynolds number and magnetic Reynolds number, are modelled through
five non-linear terms that represent known physical processes and are
strongly constrained by symmetry properties and dimensional
considerations.

The cooperation of these linear and non-linear effects explains,
without any fine tuning, the development of statistically steady,
anisotropic turbulent stresses in the shearing sheet, a local
representation of a differentially rotating disc, in agreement with
numerical simulations.  It reproduces other results seen in numerical
studies: that purely hydrodynamic turbulence is not sustained in a
flow that adequately satisfies Rayleigh's stability criterion; that
the shear stress in magnetorotational turbulence in non-Keplerian
discs is non-linearly related to the shear rate, at fixed angular
velocity; and that the shearing epicyclic motions found in warped
discs are damped by the turbulence at a slower rate than predicted on
the basis of an isotropic viscosity.  It also makes predictions for
future simulations, such as the decay of the two-dimensional wave and
the possibility of sustained turbulence in situations where the
angular velocity increases outwards.  The model has good formal
properties, being typically hyperbolic and therefore `causal',
guaranteeing the realizability of the stress tensors, and accounting
satisfactorily for energy conservation.

Only the linear part of the model is derived directly from the MHD
equations, and therefore the predictions of the model cannot be
regarded as rigorous deductions.  In particular, the non-linear
hydrodynamic stability of circular Keplerian flow remains unproven.
Nevertheless, even without an accurate calibration of its parameters,
the model almost certainly performs more faithfully than the alpha
viscosity model in a variety of circumstances, and therefore
represents an advance over previous work.

The general idea of seeking a model that connects the mean turbulent
stress to changes in the mean velocity gradient can be related to one
of the classical problems of continuum mechanics, which is to find the
constitutive relation that characterizes a particular substance and
allows the stress in a body to be related to its deformation history.
From this perspective, the present model is reminiscent of certain
classes of viscoelastic fluids.  In fact, the coexistence of the
Reynolds and Maxwell stresses, which are influenced by the mean flow
in different ways, suggests a kind of composite viscoelastic material.
The reason that a constitutive relation might be thought to exist at
all for magnetorotational turbulence is that the physics of the
magnetorotational instability is local, suggesting that the mean
stress in a volume comparable to $H^3$ should indeed depend only on
the recent deformation history of that parcel of fluid.

In recent years, numerical simulations of accretion flows subject to
the magnetorotational instability have been taken beyond the confines
of the shearing box to study the effects of cylindrical geometry, the
flow across the marginally stable orbit around a black hole, or the
boundary layer between a star and a disc (e.g. Armitage 1998; Hawley
2000; Hawley, Balbus \& Stone 2001; Armitage 2002).  These advances
are to be welcomed, and represent significant computational
achievements.  However, I believe there is still, and will continue to
exist, a need for a description such as the one developed in this
paper.  This is not just because global studies of realistic thin
discs are well beyond the current capabilities of direct numerical
simulations, but also because such a model will allow a variety of
methods, analytical and less direct numerical ones, to continue to be
used in studying accretion disc dynamics while taking seriously the
nature of magnetorotational turbulence.  It should be particularly
useful in understanding the dynamics of warped, eccentric and tidally
distorted discs, non-Keplerian accretion flows close to black holes,
and a variety of time-dependent accretion phenomena.

\section*{Acknowledgments}

I thank Charles Gammie and Jim Pringle for helpful discussions.  I
acknowledge the support of the Royal Society through a University
Research Fellowship.

\appendix

\section{Realizability}

The realizability constraint requires that the three eigenvalues of
$R_{ij}$ and the three eigenvalues of $M_{ij}$ remain non-negative
under the evolutionary model.  A full investigation of this issue
would require a consideration of a large number of degenerate cases.
For simplicity, I demonstrate here only the weaker result that no one
of the eigenvalues can become negative while the others remain
positive.

It is therefore assumed that, in the initial condition, $R_{ij}$ and
$M_{ij}$ are positive definite tensors.  If either $R_{ij}$ or
$M_{ij}$ were to develop a negative eigenvalue in the subsequent
evolution, when the first such eigenvalue passed through zero the
traces $R$ and $M$ (which are the sums of the eigenvalues) would still
be positive.  Therefore one may take $R,M>0$ below, without loss of
generality.

Define the quadratic forms $P=R_{ij}X_iX_j$ and $Q=M_{ij}Y_iY_j$,
where $X_i$ and $Y_i$ are differentiable vector fields advected
according to the time-reversible equations
\begin{equation}
  (\partial_t+u_j\partial_j)X_i-X_j\partial_iu_j+
  2\epsilon_{ikl}\Omega_kX_l=0,
\end{equation}
\begin{equation}
  (\partial_t+u_j\partial_j)Y_i+Y_j\partial_iu_j=0.
\end{equation}
Then
\begin{eqnarray}
  \lefteqn{{{{\rm D}P}\over{{\rm D}t}}=-(C_1+C_2)L^{-1}R^{1/2}P+
  {\textstyle{{1}\over{3}}}C_2L^{-1}R^{3/2}X_iX_i}&\nonumber\\
  &&+C_3L^{-1}M^{1/2}M_{ij}X_iX_j-C_4L^{-1}R^{-1/2}MP,
  \label{dpdt}
\end{eqnarray}
\begin{equation}
  {{{\rm D}Q}\over{{\rm D}t}}=C_4L^{-1}R^{-1/2}MR_{ij}Y_iY_j-
  (C_3+C_5)L^{-1}M^{1/2}Q,
  \label{dqdt}
\end{equation}
where ${\rm D}/{\rm D}t=\partial_t+u_i\partial_i$ is the Lagrangian
time-derivative.

In the initial condition, $P>0$ everywhere for all $X_i$ and $Q>0$
everywhere for all $Y_i$.  According to equation (\ref{dpdt}), ${\rm
  D}P/{\rm D}t>0$ whenever $P=0$ and $M_{ij}$ is positive definite,
and therefore $P$ cannot become negative.  According to equation
(\ref{dqdt}), ${\rm D}Q/{\rm D}t>0$ whenever $Q=0$ and $R_{ij}$ is
positive definite, and therefore $Q$ cannot become negative.

\label{lastpage}


\begin{thebibliography}{}
  
\bibitem[]{ABL96} Abramowicz M., Brandenburg A., Lasota J.-P., 1996,
  MNRAS, 281 L21
  
\bibitem[]{A78} Acheson D. J., 1978, Phil. Trans. R. Soc. London A,
  289, 459

\bibitem[]{A98} Armitage P. J., 1998, ApJ, 501, L189

\bibitem[]{A02} Armitage P. J., 2002, MNRAS, 330, 895

\bibitem[]{B95} Balbus S. A., 1995, ApJ, 453, 380

\bibitem[]{BH91} Balbus S. A., Hawley J. F., 1991, ApJ, 376, 214

\bibitem[]{BH92} Balbus S. A., Hawley J. F., 1992, ApJ, 400, 610
  
\bibitem[]{BH98} Balbus S. A., Hawley J. F., 1998, Rev. Mod. Phys.,
  70, 1
  
\bibitem[]{BT01} Balbus S. A., Terquem C., 2001, ApJ, 552, 235

\bibitem[]{B53} Batchelor G. K., 1953, The Theory of Homogeneous
  Turbulence, Cambridge Univ. Press, Cambridge
  
\bibitem[]{BNST95} Brandenburg A., Nordlund \AA., Stein R. F.,
  Torkelsson U., 1995, ApJ, 446, 741
  
\bibitem[]{B85} Burke W. L., 1985, Applied Differential Geometry,
  Cambridge Univ. Press, Cambridge

\bibitem[]{C60} Chandrasekhar S., 1960, Proc. Natl Acad. Sci., 46, 253

\bibitem[]{FT95} Foglizzo T., Tagger M., 1995, A\&A, 301, 293

\bibitem[]{F69} Fricke K., 1969, A\&A, 1, 388

\bibitem[]{GL65} Goldreich P., Lynden-Bell D., 1965, MNRAS, 130, 125
  
\bibitem[]{GT79} Goldreich P., Tremaine S., 1979, ApJ, 233, 857
  
\bibitem[]{H00} Hawley J. F., 2000, ApJ, 528, 462
  
\bibitem[]{H01} Hawley J. F., 2001, ApJ, 554, 534

\bibitem[]{HGB95} Hawley J. F., Gammie C. F., Balbus S. A., 1995, ApJ,
  440, 742
  
\bibitem[]{HGB96} Hawley J. F., Gammie C. F., Balbus S. A., 1996, ApJ,
  464, 690

\bibitem[]{HBS01} Hawley J. F., Balbus S. A., Stone J. M., 2001, ApJ,
  554, L49

\bibitem[]{HBW99} Hawley J. F., Balbus S. A., Winters W. F., 1999,
  ApJ, 518, 394

\bibitem[]{JGK01} Ji H., Goodman J., Kageyama A., 2001, MNRAS, 325, L1
  
\bibitem[]{K78} Kato S., 1978, MNRAS, 185, 629

\bibitem[]{K94} Kato S., 1994, PASJ, 46, 589

\bibitem[]{KI94} Kato S., Inagaki S., 1994, PASJ, 46, 289

\bibitem[]{KY93} Kato S., Yoshizawa A., 1993, PASJ, 45, 103
  
\bibitem[]{KY95} Kato S., Yoshizawa A., 1995, PASJ, 47, 629  

\bibitem[]{KY97} Kato S., Yoshizawa A., 1997, PASJ, 49, 213
  
\bibitem[]{LLR75} Launder B. E., Reece G. J., Rodi W., 1975, J. Fluid
  Mech. 68, 537
  
\bibitem[]{LPP94} Lyubarskij Y. E., Postnov K. A., Prokhorov M. E.,
  1994, MNRAS, 266, 583
  
\bibitem[]{MS00} Miller K. A., Stone J. M., 2000, ApJ, 534, 398
  
\bibitem[]{O01} Ogilvie G. I., 2001, MNRAS, 325, 231

\bibitem[]{OP96} Ogilvie G. I., Pringle J. E., 1996, MNRAS, 279, 152
 
\bibitem[]{OP02} Ogilvie G. I., Proctor M. R. E., 2002, J. Fluid
  Mech., submitted
 
\bibitem[]{PP83} Papaloizou J. C. B., Pringle J. E., 1983, MNRAS, 202,
  1181
 
\bibitem[]{PS92} Papaloizou J. C. B., Szuszkiewicz E., 1992, Geophys.
  Astrophys. Fluid Dyn., 66, 223
 
\bibitem[]{P96} Pumir A., 1996, Phys. Fluids, 8, 3112
  
\bibitem[]{SNHA96} Schramkowski G. P., van Niekerk E. C. M., Hoyng P.,
  Achterberg A., 1996, A\&A 315, 638

\bibitem[]{SS73} Shakura N. I., Sunyaev R. A., 1973, A\&A, 24, 337
  
\bibitem[]{SJ00} Sj\"ogren T., Johansson A. V., 2000, Phys. Fluids,
  12, 1554

\bibitem[]{S91} Speziale C. G., 1991, Annu. Rev. Fluid Mech., 23, 107

\bibitem[]{SP02} Steinacker A., Papaloizou J. C. B., 2002, ApJ, 571, 413

\bibitem[]{SHGB95} Stone J. M., Hawley J. F., Gammie C. F., Balbus S.
  A., 1996, ApJ, 463, 656
  
\bibitem[]{TP96} Terquem C., Papaloizou J. C. B., 1996, MNRAS, 279,
  767
  
\bibitem[]{TOBPNS00} Torkelsson U., Ogilvie G. I., Brandenburg A.,
  Pringle J. E., Nordlund \AA., Stein R. F., 2000, MNRAS, 318, 47
  
\bibitem[]{TP92} Tout C. A., Pringle J. E., 1992, MNRAS, 259, 604
 
\bibitem[]{V59} Velikhov E. P., 1959, Sov. Phys. JETP, 9, 995
  
\end{thebibliography}
\end{document}